\documentclass[journal,onecolumn,font=12]{IEEEtran}
 
\ifCLASSINFOpdf
   \usepackage[pdftex]{graphicx}
\else
   \usepackage[dvips]{graphicx}
\fi

\usepackage[cmex10]{amsmath}

\usepackage{amssymb}
\usepackage{color}
\usepackage{bbm}

\allowdisplaybreaks

\usepackage{array} 

\usepackage{cite}
\usepackage{url}
\usepackage{epstopdf,tikz}

\newtheorem{theorem}{Theorem}
\newtheorem{definition}{Definition}
\newtheorem{lemma}{Lemma}
\newtheorem{proposition}{Proposition}
\newtheorem{corollary}{Corollary}
\newtheorem{remark}{Remark}

\newcommand{\TV}{\text{TV}}

\newcommand{\test}{\text{test}}

\newcommand{\T}{\text{T}}
\newcommand{\Lo}{\text{L}}
\newcommand{\Hi}{\text{H}}
\newcommand{\D}{\text{D}}
\newcommand{\V}{\text{V}}
\newcommand{\w}{\text{w}}

\def\bx{{x}^n}

\DeclareMathOperator*{\argmax}{argmax}

\begin{document}

%
\title{Covert Communication Over a Compound Channel}


\author{Sadaf Salehkalaibar, Mohammad Hossein Yassaee, Vincent Y. F. Tan
	\thanks{Sadaf  Salehkalaibar is  with the Department of Electrical and Computer Engineering, College of Engineering, University of Tehran, Tehran, Iran (e-mail:  {s.saleh@ut.ac.ir})}
	\thanks{Mohammad Hossein Yassaee is with  Institute for Research in Fundamental Sciences (IPM), Tehran, Iran  (e-mail: yassaee@ipm.ir).}
	\thanks{Vincent Y. F. Tan is  with Department of Electrical and Computer Engineering and the Department of Mathematics, National University of Singapore, 
		(e-mail:  {vtan@nus.edu.sg}).}
	\thanks{Parts of the material in this paper will be presented at the \emph{IEEE International Symposium on Information Theory (ISIT), Paris, France, 2019.}}
}

\maketitle

\begin{abstract}
In this paper, we consider   fundamental communication limits over a compound channel. Covert communication in the information-theoretic context has been primarily concerned with fundamental limits when the transmitter wishes to communicate to  legitimate receiver(s) while ensuring that the communication is not detected by an adversary. This paper, however, considers an alternative, and no less important, setting in which the object to be masked is the state of the compound channel. Such a communication model has applications in the prevention of malicious parties seeking to jam the communication signal when, for example, the signal-to-noise ratio of a wireless channel is found to be low. Our main contribution is the  establishment of  bounds on the throughput-key region when the covertness constraint is defined in terms of the total variation distance. In addition, for the scenario in which the key length is infinite, we provide a sufficient condition for when the bounds coincide for the scaling of the throughput, which follows the square-root law. Numerical examples, including that of a Gaussian channel, are provided to illustrate our results. 
\end{abstract}

\IEEEpeerreviewmaketitle

\section{Introduction}


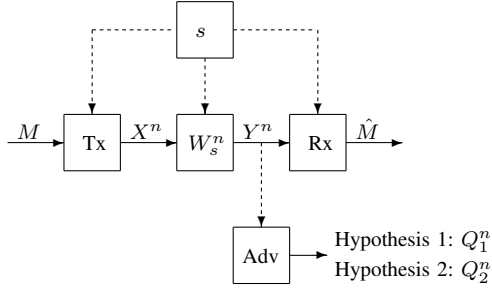
\begin{figure}[t]
	\centering
	\setlength{\unitlength}{.25mm}
	\begin{picture}(210,150)
	\put(30,60){\line(1,0){30}}
	\put(30,60){\line(0,1){30}}
	\put(30,90){\line(1,0){30}}
	\put(60,60){\line(0,1){30}}
	
	\put(90,60){\line(1,0){30}}
	\put(90,60){\line(0,1){30}}
	\put(90,90){\line(1,0){30}}
	\put(120,60){\line(0,1){30}}
	
	\put(150,60){\line(1,0){30}}
	\put(150,60){\line(0,1){30}}
	\put(150,90){\line(1,0){30}}
	\put(180,60){\line(0,1){30}}
	
	\put(90,120){\line(1,0){30}}
	\put(90,120){\line(0,1){30}}
	\put(90,150){\line(1,0){30}}
	\put(120,120){\line(0,1){30}}
	
	\put(120,00){\line(1,0){30}}
	\put(120,00){\line(0,1){30}}
	\put(120,30){\line(1,0){30}}
	\put(150,00){\line(0,1){30}}
	
	\put(00,75){\vector(1,0){30}}
	\put(60,75){\vector(1,0){30}}
	\put(120,75){\vector(1,0){30}}
	\put(180,75){\vector(1,0){30}}
	
	\multiput(45,135)(4,0){11}{\line(1,0){2}}
	\multiput(45,135)(0,-4){11}{\line(0,-1){2}}
	\multiput(120,135)(4,0){11}{\line(1,0){2}}
	\multiput(165,135)(0,-4){11}{\line(0,-1){2}}
	\multiput(105,120)(0,-4){8}{\line(0,-1){2}}
	\put(105,94){\vector(0,-1){4}}
	\put(45,94){\vector(0,-1){4}}
	\put(165,94){\vector(0,-1){4}}
	\multiput(135,75)(0,-4){11}{\line(0,-1){2}}
	\put(135,34){\vector(0,-1){4}}
	\put(150,15){\vector(1,0){20}}
	\put(174,20){\footnotesize Hypothesis 1: $Q_1^n$}
	\put(174,04){\footnotesize Hypothesis 2: $Q_2^n$}
	
	\put(40,71){\footnotesize $\mbox{Tx}$}
	\put(160,71){\footnotesize $\mbox{Rx}$}
	\put(100,131){\footnotesize $s$}
	\put(96,71){\footnotesize $W_s^n$}
	\put(125,11){\footnotesize $\mbox{Adv}$}
	\put(5,77){\footnotesize $M$}
	\put(65,77){\footnotesize $X^n$}
	\put(125,77){\footnotesize $Y^n$}
	\put(185,77){\footnotesize $\hat{M}$}
	\end{picture}
	\caption{Covert communication over a compound DMC.}
	\label{figure0}
\end{figure}

In modern applications that involve security considerations, it is not only of paramount importance that the message to be transmitted is hidden from malicious parties, but the very act of transmitting a message  should be concealed from such parties. This is the study of {\em covert communication} and has been extensively studied in the information theory community in recent years~\cite{Bash,Ligong,Bloch}. Here, the transmitter feeds a specific symbol, called the {\em off-symbol}, to the channel when it decides not to send a message, signifying that the transmitter is in the \emph{off-transmission state}.  When the transmitter decides to send a {\em bona fide} message, the channel, together with the codeword associated to the message, induce a certain output distribution that is different from that when a message is not sent. One desires to design codes such that distinguishing between these two output distributions is difficult in the sense that the failure probability is high. The problem of covert communication over a channel with an additive white Gaussian noise (i.e., an AWGN channel) \cite{Bash},  a discrete memoryless channel (DMC) \cite{Ligong,Bloch}, a classical-quantum channel \cite{Ligong2}, a channel with state \cite{Lee}, in the presence of an adversarial jammer \cite{Jaggi2},  and from the second-order perspective \cite{Bloch2,Mehrdad} have been considered. It is shown that the maximum amount of information that can be sent scales with the \emph{square root} of the transmission blocklength; this is known as the {\em square-root law}. The extension to multi-user setups has been considered for multiple-access channels \cite{Bloch3}, broadcast channels \cite{Vincent} and relay channels \cite{Ligong4}.   


In this work, we consider covert communication over a compound channel with two states $s\in\{1,2\}$ (see Fig.~\ref{figure0}). In a compound channel~\cite{blackwell, dobrushin, wolfowitz}, the channel law $W_s$ depends on the state which remains fixed during the transmission (see Section~7.2 of~\cite{ElGamal}). The channel state is known at the transmitter and the receiver. The goal of the communication is to reliably send a message to the receiver subject to a covertness constraint. We define the covertness in this unique setting as follows. An adversary who is observing the channel output cannot  reliably distinguish the channel state of communication. The motivation comes from the fact that if the adversary determines the channel state, then there may be a possibility that it interferes with the communication by consuming a small amount of power. For example, consider a wireless channel with fading and for simplicity, assume that the fading takes on one of two states in $\{\mbox{``high SNR''}, \mbox{``low SNR''}\}$. If a malicious party can deduce that the channel is in the ``low SNR'' state, then it knows that it is easier to interfere with the communication, possibly by injecting more noise to further disrupt the system by  sending jamming signals with a small power consumption. Motivated by this scenario, we consider covert communication over a compound channel. Instead of focusing solely on covertness being measured in terms of the ability of an observer at the receiver being able to distinguish between the absence or presence of message transmission, here we also quantify covertness as the ability of an adversary in distinguishing between the two channel states. That is, we quantify covertness in terms of the probability of success of an adversary in a binary hypothesis test that seeks to distinguish between $s\in \{1,2\}$.  We wish to design coding schemes such that this  probability of success  is suitably upper bounded by a small constant, meaning that the adversary's test is unreliable. Our main contribution is in the establishment of bounds that characterize the optimal tradeoff between the throughput for deniable (i.e., the adversary cannot learn whether communication is occurring or what the channel state is) and reliable communication as well as the length of the shared key between the transmitter and receiver. In \cite{Jaggi}, covert communication over compound binary symmetric channels  was  studied but the setting is different from ours. Specifically, the best throughput to deniably (i.e., the adversary cannot learn whether communication is occurring) and reliably communicate was established, whereas our main contribution is to establish bounds on the throughput under the constraint that the adversary cannot learn the channel state. 



We consider two covertness criteria for this setup. 
\begin{itemize}
\item For the first criterion, a specific distribution $Q_0$ over the channel output alphabet $\mathcal{Y}$ is fixed. This is similar to  traditional works on covert communication such as \cite{Ligong,Bloch} in which $Q_0$ is the distribution that is induced from sending the off-symbol at the input of the channel. For a given blocklength $n$, the covertness is measured by the  Kullback-Leibler (KL)-divergence of the channel output distribution, denoted as $Q_s^n$, with $Q_0^{\times n}$, which denotes the $n$-fold product of $Q_0$. In order to fix such a distribution $Q_0$, it is assumed that there exists two distributions $P_1$ and $P_2$ over the channel input alphabet $\mathcal{X}$ such that they induce the \emph{same} channel output marginal when they are fed into channels $W_1$ and $W_2$, respectively and the distribution $Q_0$ denotes this common marginal. Notice the difference of definition of $Q_0$ in the compound setup with that of a DMC in \cite{Ligong,Bloch} where $Q_0$ denotes the channel output marginal when the off-symbol is fed into the channel. Here, $Q_0$ serves to \emph{mask} the channel state and it does not necessarily correspond to the output marginal of a specific symbol in the input alphabet. Thus, the off-symbol which indicates that the transmitter is in the off-transmission state, does not necessarily exist. Even though an adversary who is observing the channel output may not be able to infer the state of the communication, he can infer whether the transmitter is sending a message.  For this covertness criterion, assuming a shared key of infinite length between the transmitter and receiver, we provide the optimal  throughput. 
\item
The second covertness criterion, as we shall see based on the motivation above, is more pertinent to the compound channel. It is defined to be the total variation distance between $Q_1^n$ and $Q_2^n$, the two output marginal distributions when the state of the channel is $ s=1$ or $s=2$.  Here, we assume that a shared key with bounded length is available between the transmitter and receiver. We provide inner and outer bounds on the throughput-key region and show that the square root law \cite{Bash} also holds in this setting. The bounds match for a special case and thus for this case, we have characterized the optimal transmission rate given a key with a certain length. The achievable scheme uses two codebooks (each of which is used  for communication given a  channel state) and a maximum likelihood (ML) decoder  at the receiver. The analysis of this scheme consists of three steps. The first step is to show that the average error probability of the scheme when averaged over all codebooks, can be appropriately upper bounded. The second step is to provide a single-letter characterization of the covertness metric. The third step is to show that there exists a deterministic codebook that satisfies the desired constraints on the maximum error probability and also the covertness criterion. The proof of outer (converse) bound is challenging since we need to analyze the total variation distance between two {\em non-product} distributions $Q_1^n$ and $Q_2^n$. The single-letter characterization of the covertness metric requires relating the total variation distance to a hypothesis testing problem and using various bounds (Berry-Esseen)  to control the false alarm and missed detection probabilities of this test. 
\end{itemize}

We conclude this section with a summary of the main contributions, introduction of notation, and the organization of the paper.

\subsection{Contributions}

The main contributions of this paper are as follows:
\begin{itemize}
	\item The problem of covert communication  for a compound DMC is formulated. 
	\item For the case in which covertness is measured in terms of the total variation distance between $Q_1^n$ and $Q_2^n$, we provide an inner bound on the the optimal transmission throughput-shared key region (Theorem~\ref{key-thm}).  The proof of this theorem involves three steps which are discussed in Lemmas \ref{lem-ach}--\ref{lem4}.
	\item If we assume that the key is sufficiently long and the error probability of decoding the message at the legitimate receiver is vanishing, we are able to  upper bound the optimal transmission throughput for the same covertness criterion (Theorem~\ref{thm-upper}). It is shown that the upper and lower bounds match for a special case (Theorem~\ref{opt-thm}). 
	\item Examples are provided to compare the lower and upper bounds. In particular, a Gaussian setup is also considered where we identify parameter regimes in which the conditions of the optimality result in Theorem~\ref{opt-thm} are satisfied. The proof of converse in this theorem needs to be adapted to the Gaussian model since the output alphabet of the Gaussian channel is continuous. 
	\end{itemize}

\subsection{Notation}

We mostly use the notation of \cite{ElGamal}. Random variables are shown by capital letters, e.g., $X$, $Y$ and their realizations by lower-case letters, e.g., $x$, $y$. The alphabets of random variables are denoted by script symbols such as $\mathcal{X}$, $\mathcal{Y}$. The sequence of random variables $(X_i,\ldots,X_j)$ and its realizations $(x_i,\ldots,x_j)$ are abbreviated as  $X_i^j$ and $x_i^j$ respectively. We use $X^j$ and $x^j$ instead of $X_1^j$ and $x_1^j$. The empirical distribution of sequence $x^n$ is also known as its \emph{type}. 

The probability mass function (pmf) of a discrete random variable $X$ defined over the channel input alphabet $\mathcal{X}$ denoted using the letter $P$. The pmf of a discrete random variable $Y$ defined over the channel output alphabet $\mathcal{Y}$ is denoted using the letter $Q$. The distributions of the  sequences of random variables $X^n$ and $Y^n$ are denoted by $P^n$ and $Q^n$ respectively.    The notation $P^{\times n}$ denotes the $n$-fold product distribution, i.e., for every $x^n\in\mathcal{X}^n$, we have
\begin{align}
P^{\times n}(x^n):=\prod_{i=1}^n P(x_i).
\end{align}
The expectation and variance operators are written as $\mathbb{E}[.]$ and $\mathbb{V}[.]$, respectively. The notation $\mathbbm{1}\left\{ . \right\}$ denotes the indicator function.  The probability of an event $\mathcal{E}\subseteq \mathcal{X}$ is denoted by $\mathbb{P}(\mathcal{E})$. 

The KL-divergence, chi-squared distance and total variation distance between two distributions $P_1$ and $P_2$ are denoted by $D(P_1\| P_2)$, $\chi_2(P_1\| P_2)$ and $d_{\TV}(P_1,P_2)$, respectively. The Bhattacharyya distance between two distributions $P_1$ and $P_2$ is denoted by $F(P_1,P_2)$ and is defined as $F(P_1,P_2)\triangleq \sum_x\sqrt{P_1(x)P_2(x)}$.

The Hamming weight of sequence $x^n$ is denoted by $\w(x^n)$. The binary entropy function is shown by $h_{\text{b}}(.)$. The binary symmetric channel with parameter $\epsilon$ is denoted by $\text{BSC}(\epsilon)$. The Bernoulli distribution, i.e., one that is defined over $\{0,1\}$, is denoted by $\text{Bern}(.)$. The cumulative distribution function of a standard normal distribution is denoted by $\Phi(.)$. All logarithms are with respect to the base 2. The exponential function is written as $\exp(.)$.

\subsection{Organization}
The remainder of the paper is organized as follows. Section~\ref{covert-def} describes the covert communication setup over a compound channel. Section~\ref{sec:enforce} provides the lower and upper bounds on the optimal transmission throughput and some examples to discuss the bounds. The paper is concluded in Section~\ref{sec:conclusion} and by technical appendices.

\section{Covert Communication over a Compound Channel}\label{covert-def}

\subsection{System Model}\label{sys-model1}

Consider covert communication over a compound DMC  with input and output alphabets $\mathcal{X}$ and $\mathcal{Y}$, see Fig.~\ref{figure0}. The compound channel consists of two channel pmfs $W_s( \cdot |\cdot ), s\in\{1,2\}$ where the channel state $s$ is available at both the transmitter and the receiver. The message set is shown by $\mathcal{M}$. The transmitter and receiver share a secret key $K\in \mathcal{K}$. The transmitter  sends an $n$-length input $X^n$ over the channel using a state-dependent encoding function $f_s^{(n)}\colon\mathcal{M}\times \mathcal{K}\to \mathcal{X}^n$ where  
\begin{align}
X^n &= f_s^{(n)}(M,K).
\end{align}
The receiver uses a decoding function $g_s^{(n)}\colon \mathcal{Y}^n\times \mathcal{K}\to \mathcal{M}$ which maps the channel output to an estimated message as follows:
\begin{align}
\widehat{M} &= g_s^{(n)}(Y^n,K).
\end{align}

In the conventional covert communication over a single-state point-to-point channel \cite{Ligong,Bloch}, a fixed distribution which is denoted by $Q_0$, is chosen to represent the  single-letter channel output marginal when the transmitter sends the off-symbol. This symbol is fed to the channel when the transmitter decides not to send a message. The covertness is measured by the ``distance'' between the channel output marginal when a message is sent and the distribution $Q_0^{\times n}$.

The goal of covert communication over a compound channel is primarily to conceal the channel state. It might be possible to cover message transmission such that an adversary observing the channel output does not know whether an off-symbol or a message is sent.  In the following, we consider the different cases where a distribution $Q_0$ can be fixed for the definition of covert communication over a compound channel.

\underline{\textit{Case 1 (Only concealing the channel state)}}: In this case, the channel state is masked. However, the message transmission might be revealed to an adversary since an off-symbol does not exist. 
Specifically, there exist two non-deterministic distributions\footnote{A {\em deterministic distribution} $P$ is one in which there exists a symbol $x^*$ such that $P(x^*)=1$. A {\em non-deterministic distribution} is one that is not deterministic.} $P_1$ and $P_2$ such that 
\begin{align}
\sum_x P_1(x)W_1(y|x)=\sum_xP_2(x)W_2(y|x),\; \forall y\in\mathcal{Y}.\label{covert-equation}
\end{align} 
In this case, we define $Q_0$ to represent the above common channel output marginal between the two states. Here, an adversary who is observing the channel output, cannot infer the channel state. Thus, the distribution $Q_0$ serves to {\em mask} the state of the communication. Here, there does not necessarily exist an input symbol whose channel output marginal is as \eqref{covert-equation}. Thus, the adversary who is observing the channel output may infer whether the transmitter is sending a message.

An example of this case of covert communication is discussed in the following. Suppose that $\mathcal{X}=\mathcal{Y}=\{0,1\}$, where $W_1(.|.)$ and $W_2(.|.)$ are $\text{BSC}\left(\frac{1}{4}\right)$ and $\text{BSC}\left(\frac{1}{3}\right)$, respectively. The  non-deterministic input distributions $P_1,P_2\sim \text{Bern}\left(\frac{1}{2}\right)$ satisfy Condition \eqref{covert-equation} and $Q_0(0)=Q_0(1)=1/2$. Clearly, the distribution $Q_0$ is not induced by any of the input symbols at both channel states.\\

\underline{\textit{Case 2 (Concealing channel state and message transmission)}}:  In this case, condition \eqref{covert-equation} is satisfied only if $P_1$ and $P_2$ are deterministic distributions. Thus, there exist  symbols
``$0$''  and ``$0'$" such that
\begin{align}
W_1(y|0)=W_2(y|0'),\qquad \forall y\in\mathcal{Y}.
\end{align}
Without loss of generality, we can suppose that the symbols $0$ and $0'$ are equal, because the transmitter has access to the state\footnote{In fact, if $0\neq 0'$, we can permute the input symbols for the state $s=2$, such that the $0'$ is renamed to $0$.}.
Here, we again call the symbol $0$ as \emph{off-symbol}. Let $Q_0$ denotes the above common channel output marginal between the two channel states. The distribution $Q_0$ masks the channel state and also represents the scenario in which the transmitter does not send a message at both channel states. 

\begin{figure}[t]
	\centering
	\includegraphics[scale=1.1]{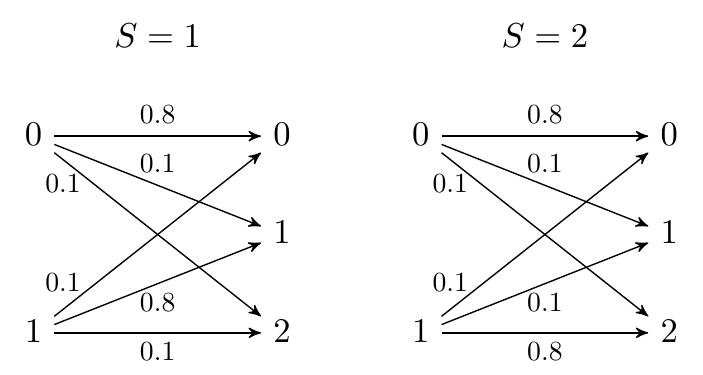}
	\caption{Example for Case 2.}
	\label{figure2}
\end{figure}

An example of this case is given in Fig.~\ref{figure2} where we assume that $\mathcal{X}=\{0,1\}$ and $\mathcal{Y}=\{0,1,2\}$. Notice that the unique distributions satisfying \eqref{covert-equation} are given by $P_1(0)=P_2(0)=1$ and $P_1(1)=P_2(1)=0$.
Thus, the symbol ``0'' can be regarded as the off-symbol.\\

\underline{\textit{Case 3 (Concealing channel state and message transmission partially)}}: Condition  \eqref{covert-equation} is satisfied only if $P_1$ is deterministic while $P_2$ can be non-deterministic. In this case, we have:
\begin{align}
W_1(y|0) = \sum_xP_2(x)W_2(y|x),\qquad \forall y\in\mathcal{Y}.
\end{align}
Here, $Q_0$ denotes the above common channel output marginal. Besides the fact that $Q_0$ masks the channel state, it also represents the fact that the transmitter has no input to the channel when $S=1$. A symmetric case can be considered when $P_2$ is deterministic and $P_1$ is non-deterministic.\\

\underline{\textit{Case 4 (Revealing channel state and message transmission)}}: There does not exist any pair of distributions $P_1$ and $P_2$ such that Condition~\eqref{covert-equation} is satisfied. In this case,  covert communication over the compound channel is not feasible.



\subsection{Covertness Criteria}

As discussed in the above Cases 1--3, a distribution $Q_0$ over $\mathcal{Y}$ can be fixed. As such we can define the covert rate in a number of different ways depending on the covertness criterion. Two different covertness criteria are considered.

\begin{definition}\label{def-main} An $(n,|\mathcal{M}|, |\mathcal{K}|)$-code consists of a message set $\mathcal{M}$, a key set $\mathcal{K}$, two encoding functions $f_s^{(n)}:\mathcal{M}\times \mathcal{K}\to\mathcal{X}^n$, and two decoding functions $g_s^{(n)}:\mathcal{Y}^n\times \mathcal{K}\to\mathcal{M}$ for $s\in\{1,2\}$.
For each state $s$, the code induces the following joint pmf on the random variables $(M,K,X^n,Y^n,\widehat{M})$\footnote{The adversary knows the distribution which is used for generating the codebook but does not know the chosen codebook by the transmitter and receiver.}:
\begin{align}
P_{MKX^nY^n\widehat{M}}(m,k,x^n,y^n,\hat{m})\triangleq \frac{1}{|\mathcal{M}|\cdot |\mathcal{K}|} W_s^n(y^n|x^n)\cdot \mathbbm{1}\left\{ f_s^{(n)}(m,k)=x^n \right\}\cdot \mathbbm{1}\left\{ g_s^{(n)}(y^n,k)=\hat{m} \right\}.
\end{align}
We denote the channel output marginal on $\mathcal{Y}^n$ by $Q_s^n$.
Define:
	\begin{IEEEeqnarray}{rCl}
	P_{\mathrm{e}}(s,k,m)\triangleq \mathbb{P}\big(\widehat{M}\neq M|S=s, K=k, M=m\big),
\end{IEEEeqnarray}
and
\begin{IEEEeqnarray}{rCl}
	\bar{P}_{\mathrm{e}}(s,k)\triangleq \frac{1}{|\mathcal{M}|}\sum_{m=1}^{|\mathcal{M}|} \mathbb{P}\left(\widehat{M}\neq M| M=m,K=k,S=s\right),\qquad \forall s\in\{1,2\},\;k\in\mathcal{K}.\label{average-def}
\end{IEEEeqnarray}
For this code, the maximum error probability is defined as 
	\begin{IEEEeqnarray}{rCl}
		P_{\mathrm{e}} \triangleq \max_{s=1,2}\;\; \max_{k\in\mathcal{K}}\;\;\max_{m\in\mathcal{M}}\;P_{\mathrm{e}}(s,k,m).\label{Pe-def}
	\end{IEEEeqnarray}	

Two covertness criteria  are considered. The first criterion is that  the KL-divergence between the marginal distributions $Q_s^n, s=1,2$ and the \emph{masking} distribution $Q_0^{\times n}$ is sufficiently small. This criterion imposes more restriction on the concealing the channel state in the sense that not only adversary can not infer the channel state, it also can not distinguish the marginal outputs from the masking distribution. 
The second criterion relaxes this additional requirement. The second criterion requires that  the total variation distance between the channel output marginals of the two states, i.e., $d_{\TV}(Q_1^n,Q_2^n)$ to be small. Thus, adversary  can not  only infer the channel state.

 Therefore, we have two definitions:
\begin{itemize}
	\item An $(n,|\mathcal{M}|,|\mathcal{K}|,\epsilon,\delta_1,\delta_2,\zeta)_{\mathrm{KL}}$-code is an $(n,|\mathcal{M}|,|\mathcal{K}|)$-code such that the KL-divergence covertness terms are no larger than $\delta_1$ and $\delta_2$, when the state is $s=1$ and $s=2$, respectively and the maximum error probability is no larger than $\epsilon$.
	\item An $(n,|\mathcal{M}|,|\mathcal{K}|,\epsilon,\delta)_{\mathrm{TV}}$-code is an $(n,|\mathcal{M}|,|\mathcal{K}|)$-code such that the total variation distance as the covertness metric is no larger than $\delta$ and the maximum error probability is no larger than $\epsilon$.\\
	\end{itemize}\end{definition}

\begin{definition}[KL-divergence criterion] 
	For $(\epsilon,\delta_1,\delta_2, \zeta) \in\mathbb{R}_+^4$, a pair  $(L,L_K)\in\mathbb{R}_+^2$ is said to be $(\epsilon,\delta_1,\delta_2,\zeta)$-achievable if there exists a sequence of  $(n,|\mathcal{M}|, |\mathcal{K}|, \epsilon,\delta_1,\delta_2,\zeta)_{\mathrm{KL}}$-codes such that 	\begin{align}
	\liminf_{n\to \infty}\; \frac{\log |\mathcal{M}|}{n^{\zeta}}\ge L, \label{eqn:liminf_L}
	\end{align}
	\begin{equation}
	\limsup_{n\to\infty}\; \frac{\log |\mathcal{K}|}{n^{\zeta}}\leq L_K.\label{liminf_LK}
	\end{equation}
	The set of all $(\epsilon,\delta_1,\delta_2,\zeta)$-achievable  pairs $(L,L_K)$ is denoted as $\mathcal{L}_{\zeta}^*(\epsilon,\delta_1,\delta_2)$. We also define $L_{\zeta}^*(\epsilon,\delta_1,\delta_2)$ as follows:
	\begin{IEEEeqnarray}{rCl}
		L_{\zeta}^*(\epsilon,\delta_1,\delta_2)=\sup\left\{ L\colon (L,L_K)\in \mathcal{L}^*(\epsilon,\delta_1,\delta_2,\zeta)\;\text{for some}\; L_K \right\},
	\end{IEEEeqnarray}	
	and 
	\begin{IEEEeqnarray}{rCl}
		L_{\zeta}^*(0^+,\delta_1,\delta_2)=\lim_{\epsilon\to 0} L_{\zeta}^*(\epsilon,\delta_1,\delta_2).
	\end{IEEEeqnarray}
	\end{definition}



\begin{remark} In this work, we are interested in the cases of $\zeta=\frac{1}{2}$ and $\zeta=1$. When $L_{1}^*(\epsilon,\delta_1,\delta_2)>0$, a positive rate of communication is achievable.\\
	\end{remark}

\begin{definition}[Total variation distance as the covertness criterion]\label{TV-def}
 For $(\epsilon,\delta)\in\mathbb{R}_+^2$, a pair  $(L,L_K)\in\mathbb{R}_+^2$ is said to be  $(\epsilon,\delta)$-achievable if there exists a sequence of  $(n,|\mathcal{M}|,|\mathcal{K}|,\epsilon,\delta)_{\mathrm{TV}}$-codes such that \eqref{eqn:liminf_L}--\eqref{liminf_LK} hold for $\zeta=\frac{1}{2}$. The set of all $(\epsilon,\delta)$-achievable  pairs $(L,L_K)$ is denoted as  $\mathcal{L}_{\TV}^*(\epsilon,\delta)$. We also define
		\begin{IEEEeqnarray}{rCl}
			L_{\TV}^*(\epsilon,\delta)=\sup\left\{ L\colon (L,L_K)\in \mathcal{L}_{\TV}^*(\epsilon,\delta)\;\text{for some}\; L_K \right\},
		\end{IEEEeqnarray}
		and 
		\begin{align}
		L_{\TV}^*(0^+,\delta)=\lim_{\epsilon\to 0}L_{\TV}^*(\epsilon,\delta).
		\end{align}
\end{definition}

\subsection{Basic Results}
In this section, we present some basic results on the covert rates to provide more intuition on the model. For simplicity, we assume that an infinite-length key is shared between the transmitter and receiver. The covertness criterion is assumed to be the KL-divergence as in Definition~\ref{def-main}. The following proposition states that we can communicate at a positive rate in Case~1. Define:
	\begin{align}
\mathcal{P}\triangleq \left\{ (P_1,P_2)\colon \;
\text{ $(P_1,P_2)$ satisfies Condition~\eqref{covert-equation} } \right\}.
\end{align}
\begin{proposition}\label{prop1} Assuming that Case~1 holds and an infinite-length key is shared between the transmitter and receiver, we have: 
	\begin{align}
	L_1^*(0^+,\delta_1,\delta_2)=\min_{s\in\{1,2\}}\;\max\;I(P_s,W_s),\label{rate-positive-example}
	\end{align}
	where the maximum is over all distributions $(P_1,P_2)\in\mathcal{P}$.
\end{proposition} 
\begin{IEEEproof} Similar to \cite[Proposition 1]{Ligong}.
\end{IEEEproof}
An interesting example of Case~1 is the Gaussian setup which we now discuss in the following. Suppose that at each channel state $s\in\{1,2\}$, the channel is given by:
\begin{align}
Y=X+Z_s,\hspace{1cm} Z_s\sim\mathcal{N}(0,\sigma_s^2),
\end{align}
where $\sigma_1^2> \sigma_2^2$, $X,Y,Z_s\in \mathbb R$ and $X$ is independent of $Z_s$. We impose an average power constraint on the input such that $\mathbb{E}\left[|X|^2\right]\leq \mathtt{P}$ for some $\mathtt{P}>0$. In the following, we discuss the scenario in which  there exist two non-deterministic distributions $P_1$ and $P_2$ such that Condition \eqref{covert-equation} holds. Thus, a positive rate can be achieved for this example. Consider Proposition~\ref{prop1} and let the second moment of the distribution $P_s$ be denoted by $\rho_{s}$. We know that a Gaussian random variable maximizes $I(P_s,W_s)$ among all distributions of the same second moment. Thus, we get:
\begin{align}
I(P_s,W_s)\leq \frac{1}{2}\log\left(1+\frac{\rho_{s}}{\sigma_s^2}\right).
\end{align}
The distributions $P_1$ and $P_2$ satisfy \eqref{covert-equation} if the following holds:
\begin{align}
\rho_1+\sigma_1^2=\rho_2+\sigma_2^2,
\end{align}
which yields the following optimal positive rate (with $\rho_2=\mathtt{P}, \rho_1=\mathtt{P}-(\sigma_1^2-\sigma_2^2)<\mathtt{P}$):
\begin{align}
L_1^*(0^+,\delta_1,\delta_2)&= \frac{1}{2}\log\left(\frac{\mathtt{P}+\sigma_2^2}{\sigma_1^2}\right).
\end{align}
Notice the difference of this example with the Gaussian setup of \cite[Section~V]{Ligong}. In the proposed example, a positive rate can be achieved at each channel state. However,  in the Gaussian example of \cite[Section~V]{Ligong},  the maximum communication rate in the presence of covertness constraint is zero, i.e., we cannot communicate at a positive rate. As discussed in the previous section, the difference arises from the fact that the goal of covert communication over a compound channel is to mask the state, while in the conventional model, the aim is to keep the adversary oblivious about the state of transmitter,  i.e., whether or not a message is sent.  

The following proposition presents the optimal covert communication rate of Case~2 which also states that the maximum rate is zero. 

\begin{proposition}\label{prop2} Assuming that Case~2 holds and an infinite-length key is shared between the transmitter and receiver, we have
	\begin{align}
	&L_{\frac{1}{2}}^*(0^+,\delta_1,\delta_2)= \min_{s\in\{1,2\}}\; \max_{\tilde{P}_s  : \tilde{P}_s(0)=0} \sqrt{\frac{2\delta_s}{ \chi_2(\tilde{Q}_s\|Q_0)}}\cdot  D(W_s\|Q_0\big|\tilde{P}_s),\label{zero-rate-final}
	\end{align}
	where 
	$\tilde{Q}_s$ denotes the output marginal of the channel $W_s$ when the input distribution is $\tilde{P}_s$. 
\end{proposition}
\begin{IEEEproof} Similar to \cite[Theorem~2]{Ligong}.
\end{IEEEproof}
Recall the example in Fig.~\ref{figure2} which satisfies the condition of Case~2. Evaluation of \eqref{zero-rate-final} for the proposed example with $\delta_s=1$ yields $L_{\frac{1}{2}}^*(0^+ ,1,1)=1.26$.\\

To conclude this section, we briefly discuss the advantage of 
concealing only the channel state 
as compared to 
the set-up which requires the strengthen constraint that adversary cannot distinguish between the channel output and the masking distribution. Whenever Case~2 occurs, the latter setup is equivalent to protecting the message transmission from the detection by an adversary, since we have an off-symbol in Case~2. Thus, we compare the message transmission concealing setup versus concealing only the channel state.

\begin{figure}[t]
	\centering
	\includegraphics[scale=1.1]{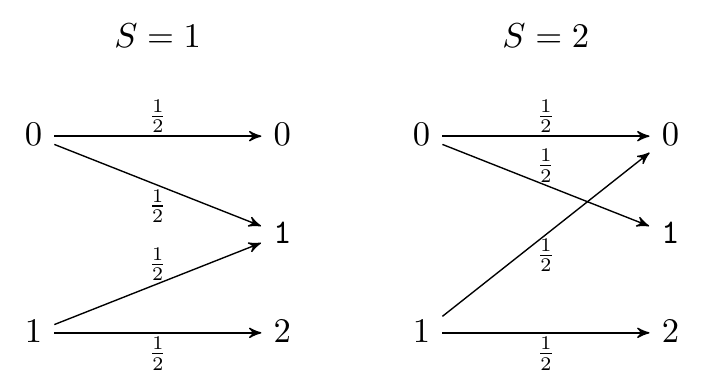}
	\caption{Example where 
	concealing the message transmission is impossible, while concealing the channel state is feasible.
	}
	\label{figure5}
\end{figure}

 Consider the setup of Fig.~\ref{figure5} which satisfies the condition of Case~2.  For this example, the KL-divergence $D(Q_s^n\|Q_0^{\times n})$ goes to infinity which implies that it is 
 not
  feasible to conceal the message transmission. 
   However, as will be discussed in Subsection~\ref{subsec:concealing}, 
   concealing the channel state under the total variation covertness criterion 
   is possible in the sense that $L^*_{\TV}(0^+,\delta)>0$.\footnote{This observation is clearly true  for the Case~1.}
\begin{remark}\label{rem2} In the following results, we assume that all the transition probabilities $W_s(.|x), s\in\{1,2\}, x\in\mathcal{X}$ are absolutely  continuous with respect to $Q_0(.)$ (if it exists). This condition ensures us that examples such as that of Fig.~\ref{figure5} are excluded and all involved information quantities are finite. If this assumption is not required, it will be explicitly mentioned in the text.
	\end{remark} 

 \section{Main Results}\label{sec:enforce} 

 In this section, we present our results when the covertness criterion is considered to be the total variation distance between channel output marginals of the two states (see Definition~\ref{TV-def}). As discussed in the previous section (see Case~1), if there exist two non-deterministic input distributions $P_1$ and $P_2$ such that Condition~\eqref{covert-equation} holds, then the optimal rate is given by Proposition~\ref{prop1} and the maximum covert communication rate is positive. Thus, we only focus on Case 2 in the following and assume that Condition~\ref{covert-equation} is satisfied only if $P_1$ and $P_2$ are deterministic input distributions.  Moreover, recall 
  ``0" is the off-symbol
 such that $ W_1(\cdot |0) = W_2(\cdot |0)\triangleq Q_0(\cdot ) $.


\subsection{An Inner Bound to $\mathcal{L}_{\TV}^*(\epsilon,\delta)$}\label{sec:optimal}

In the following, we present an inner bound to $\mathcal{L}_{\TV}^*(\epsilon,\delta)$.  The proof is given afterwards.\\
 For positive $\gamma_1,\gamma_2$ and any three  distributions $Q_0$, ${Q}_1$ and ${Q}_2$ over $\mathcal{Y}$, define:
	\begin{align}
\Omega(\gamma_1,\gamma_2;Q_0,Q_1,Q_2)&\triangleq\gamma_1^2\chi_2({Q}_1\|Q_0)+\gamma_2^2\chi_2({Q}_2\|Q_0)-2\gamma_1\gamma_2\rho({Q}_1,{Q}_2,Q_0),\label{prep1} \end{align}
where
\begin{align}
\rho({Q}_1,{Q}_2,Q_0)\!\triangleq \!\mathbb{E}_{Q_0}\left[\left(\frac{{Q}_1(Y)}{Q_0(Y)}-1\right)\cdot \left(\frac{{Q}_2(Y)}{Q_0(Y)}-1\right)\right].
\label{eqn:def_rho}
\end{align}

\begin{theorem}\label{key-thm}  An inner bound to $\mathcal{L}_{\TV}^*(\epsilon,\delta)$, is given by the union of all pairs $(L,L_K)$ such that:
	\begin{IEEEeqnarray}{rCl}
		L\leq \min_{s\in\{1,2\}}\gamma_{s}
		D(W_s \| Q_0|\bar{P}_s),\label{key-ineq1}
	\end{IEEEeqnarray}	
	and 
	\begin{IEEEeqnarray}{rCl}
		L_K \geq \max_{s\in\{1,2\}}\gamma_{s}
			D(W_s \| Q_0|\bar{P}_s)
		-\min_{s\in\{1,2\}}\gamma_{s}
			D(W_s \| Q_0|\bar{P}_s)
		,\label{key-ineq2}
	\end{IEEEeqnarray}
	where the union is over all input distributions $\bar{P}_s$ such that $\bar{P}_s(0)=0$ with their output marginals through $W_s$ are denoted by $\bar{Q}_s$ and over all  positive $  \gamma_1,\gamma_2$ such that 
	\begin{equation}
	\Omega(\gamma_1,\gamma_2;Q_0,\bar{Q}_1,\bar{Q}_2)\le \left(2\Phi^{-1}\Big(\frac{1+\delta}{2}\Big)\right)^2.
	\label{omega-ineq}
	\end{equation}
\end{theorem}
\begin{IEEEproof} The achievable scheme is sketched as follows. Two codebooks are generated where each of them is used for communicating over each channel state. The optimal decoder --namely, the ML decoder-- is used to determine the message which is sent. The analysis of error probability involves applying Shannon's achievability bound \cite[Thm~2]{Yury} to upper bound the error probability averaged over all codebooks. A main step of the analysis is the single-letter characterization of the covertness metric. Finally, it is shown that there exists a ``good'' codebook which satisfies desired properties on the maximum error probability and also the covertness criterion.

\underline{\textit{Preparations}}: Fix a large blocklength $n$, choose positive $\gamma_s$ for $s\in\{1,2\}$. Let $\bar{P}_{s}$ be any distribution such that $\bar{P}_{s}(0)=0$ and denote the output marginal through channel $W_s$ by $\bar{Q}_{s}$.
Choose positive $\gamma_s$ such that \eqref{omega-ineq} holds and let
\begin{equation}
\mu_{s,n}\triangleq \frac{\gamma_s}{\sqrt{n}}.\label{prep2}
\end{equation}
Define  the input distribution $P_{s,n}$ as 
\begin{align}
P_{s,n}(x) \triangleq \left\{\begin{array}{ll} \mu_{s,n}\bar{P}_{s}(x) & x\neq 0\\1-\mu_{s,n} & x=0\end{array}\right. .\label{step28}
\end{align}
For $s\in\{1,2\}$, let $Q_{s,n}$ denote the output marginal of the channel $W_s$ when the input distribution is $P_{s,n}$. The distribution $Q_{s,n}$ can be written as follows:
\begin{IEEEeqnarray}{rCl}
	Q_{s,n}=(1-\mu_{s,n})Q_0+\mu_{s,n}\bar{Q}_{s}.\label{Qs-def}
\end{IEEEeqnarray}
Denote the $n$-fold products of $P_{s,n}$ and $Q_{s,n}$ by $P_s^{\times n}$ and $Q_s^{\times n}$, respectively.

\underline{\textit{Codebook Generation}}:
We generate an i.i.d.\ codebook $\mathcal{C}_s =  \left\{ x_s^n(m,k)\colon m\in\mathcal{M},k\in\mathcal{K} \right\}$ according to the input distribution $P_{s,n}$ as defined in \eqref{step28}. Denote the channel output marginal by $Q_s^n$.

\underline{\textit{Single-letter Characterization of Covertness Condition}}: The following two lemmas together provide a single-letter characterization of the covertness condition.
\begin{lemma}\label{lem-ach} We have
	\begin{align} 
	\hspace{0cm}	d_{\TV}\left(Q_1^{\times n},Q_2^{\times n}\right)\leq 2\Phi\left(\frac{1}{2}\sqrt{\Omega(\gamma_1,\gamma_2;Q_0,\bar{Q}_1,\bar{Q}_2)}\right)-1+O\left(\frac{1}{\sqrt{n}}\right). \label{eqn:dTV}
	\end{align}
\end{lemma}
\begin{IEEEproof} See Appendix \ref{lem-ach-proof}.
\end{IEEEproof}

\begin{lemma}\label{lem-key} Let $\kappa$ be an arbitrary positive number. Assume that,
	\begin{align}
	\log |\mathcal{M}|+\log |\mathcal{K}| \geq (1+\kappa)\sqrt{n} \cdot \max_{s\in\{1,2\}}\gamma_s
	D(W_s \| Q_0|\bar{P}_s)
	.\label{lem-ineq1}
	\end{align}
	Then, we have
	\begin{align} 
	\hspace{0cm}	\mathbb{E}\left[d_{\TV}\left(Q_s^{ n},Q_s^{\times n}\right)\right]= O\left(\frac{1}{\sqrt{n}}\right), \label{exp:dTV}
	\end{align}
	whenever $\Omega(\gamma_1,\gamma_2;Q_0,\bar{Q}_1,\bar{Q}_2)$ satisfies \eqref{omega-ineq}.
\end{lemma}
\begin{IEEEproof} See Appendix~\ref{lem-key-proof}.
\end{IEEEproof}

\underline{\textit{Encoding and Decoding}}:
The transmitter upon observing the message $m\in\mathcal{M}$ and the key $k\in\mathcal{K}$, sends the codeword $x_s^n(m,k)$ over the channel. Given  $y^n$, the decoder based on the secret key $k$ and the state $s$ uses
the optimal ML decoder
 which chooses the message $\hat{m}$ as follows:
\begin{align}
\hat{m}=\argmax_{m\in\mathcal{M}} W^n_s(y^n|\bx_s(m,k)) 
\label{eqn:rate_con}
\end{align}


The following lemma provides an upper bound on the average error probability as defined in \eqref{average-def}.
\begin{lemma}\label{PPM-thm}   If 
	\begin{IEEEeqnarray}{rCl}
		\frac{\log |\mathcal{M}|}{\sqrt{n}}\leq \min_{s\in\{1,2\}}\sqrt{n}I(P_{s,n},W_s)-n^{-\frac{1}{6}},\label{lem4-ineq}
		\end{IEEEeqnarray}
	Then, 
	\begin{IEEEeqnarray}{rCl}
		\mathbb{E}\left[\bar{P}_{\mathrm{e}}(s,k)\right]= O\left(n^{-\frac{1}{6}}\right),\qquad \forall s\in\{1,2\},\;k\in\mathcal{K}.
		\end{IEEEeqnarray}
	
\end{lemma}
\begin{IEEEproof} See Appendix~\ref{rate-analysis}.
	\end{IEEEproof}	

\underline{\textit{Existence of a Codebook with Desired Properties}}: The following lemma proves existence of a codebook with desired properties.
\begin{lemma}\label{lem4} If for some positive $\kappa$ and $\epsilon$, constraints \eqref{lem-ineq1} and \eqref{lem4-ineq} hold, then there are exists at least one ``good'' codebook such that:
	\begin{align} 
	\hspace{0cm}	d_{\TV}\left(Q_s^{ n},Q_s^{\times n}\right)=
	o(1)
	, \label{lem4-ineq2}
	\end{align}
	and the maximum error probability as defined in \eqref{Pe-def}, satisfies:
	\begin{align}
 P_{\mathrm{e}} = o(1)
	\end{align}
	\end{lemma}
\begin{IEEEproof} See Appendix~\ref{existence2-proof}.
	\end{IEEEproof}
The proof is concluded by combining lemmas \ref{lem-ach} to \ref{lem4}. See Appendix~\ref{key-thm-proof} for details.

\end{IEEEproof}

Now, we specialize Theorem~\ref{key-thm} to the case where the length of key is sufficiently long. Also, we restrict to the case of $\mathcal{X}=\{0,1\}$ and the output distribution induced by each input symbol is denoted by
\begin{IEEEeqnarray}{rCl}
	\tilde{Q}_{1}(\cdot )\triangleq  W_1(\cdot |1),\qquad	\tilde{Q}_{2}(\cdot )\triangleq  W_2(\cdot |1).\label{output-dist-def}
\end{IEEEeqnarray}
This special case will be used later to provide an optimality result. In this case, we are able to find a closed form for the optimization problem \eqref{key-ineq1} under the constraint \eqref{omega-ineq}. The solution of this optimization is given in the following corollary.

\begin{corollary}\label{coro} Assume that the length of the key is infinite, i.e., $L_K=\infty$ and $\mathcal{X}=\{0,1\}$. Then, $L_{\TV}^*(\epsilon,\delta)$ is lower bounded as follows:
	\begin{align}
	L_{\TV}^*(\epsilon,\delta) \geq \mathsf{L},\label{L*}
	\end{align}
	where 
	
	\begin{IEEEeqnarray}{rCl}
		\mathsf{L} \triangleq  2\Phi^{-1}\left(\frac{1+\delta}{2}\right)\cdot \frac{D(\tilde{Q}_1\|Q_0) D(\tilde{Q}_2\|Q_0)}{\sqrt{\chi_2(\tilde{Q}_1\|Q_0) D(\tilde{Q}_2\|Q_0)^2-2\rho(\tilde{Q}_1,\tilde{Q}_2,Q_0) D(\tilde{Q}_1\|Q_0) D(\tilde{Q}_2\|Q_0)+\chi_2(\tilde{Q}_2\|Q_0) D(\tilde{Q}_1\|Q_0)^2}},\nonumber\\\label{L-def1}
		\end{IEEEeqnarray}
	if 
	\begin{IEEEeqnarray}{rCl}
		\rho(\tilde{Q}_1,\tilde{Q}_2,Q_0)< \min \left(\frac{\chi_2(\tilde{Q}_1\|Q_0)D(\tilde{Q}_2\|Q_0)}{D(\tilde{Q}_1\|Q_0)},\frac{\chi_2(\tilde{Q}_2\|Q_0) D(\tilde{Q}_1\|Q_0)}{D(\tilde{Q}_2\|Q_0)}\right),
		\end{IEEEeqnarray}
and 
\begin{IEEEeqnarray}{rCl}
	\mathsf{L} \triangleq 2\Phi^{-1}\left(\frac{1+\delta}{2}\right)\cdot \sqrt{\frac{\min\left(\chi_2(\tilde{Q}_1\|Q_0)D(\tilde{Q}_2\|Q_0)^2,\chi_2(\tilde{Q}_2\|Q_0) D(\tilde{Q}_1\|Q_0)^2\right)}{\chi_2(\tilde{Q}_1\|Q_0) \chi_2(\bar{Q}_2\|Q_0)-\rho(\tilde{Q}_1,\tilde{Q}_2,Q_0)^2}},
	\end{IEEEeqnarray}	
if
\begin{IEEEeqnarray}{rCl}
	 \rho(\tilde{Q}_1,\tilde{Q}_2,Q_0) \geq \min \left(\frac{\chi_2(\tilde{Q}_1\|Q_0)D(\tilde{Q}_2\|Q_0)}{D(\tilde{Q}_1\|Q_0)},\frac{\chi_2(\tilde{Q}_2\|Q_0) D(\tilde{Q}_1\|Q_0)}{D(\tilde{Q}_2\|Q_0)}\right).
	\end{IEEEeqnarray}	
	\end{corollary}
\begin{IEEEproof} See Appendix~\ref{cor-proof}.
	\end{IEEEproof}
$\\$

\subsection{An Upper Bound on $L_{\TV}^*(0^+,\delta)$ and an Optimality Result}\label{sec:upper}
 In this section, we provide an optimality result for $L_{\TV}^*(0^+,\delta)$. We consider two simplifications on the system model. First, it is assumed that the length of key is infinite. Second, we restrict to the case of $\mathcal{X}=\{0,1\}$ where the output distributions are defined in \eqref{output-dist-def}.
The following theorem presents an upper bound on $L_{\TV}^*(0^+,\delta)$. Its proof is given after the statement of Theorem~\ref{opt-thm}. Define:
\begin{IEEEeqnarray}{rCl}
	\Delta &\triangleq& \mathbb{E}_{Q_0} \bigg[\bigg( \frac{\tilde{Q}_1(Y)-\tilde{Q}_{2}(Y)}{Q_0(Y)}\bigg)^2\bigg].
\end{IEEEeqnarray}
\begin{theorem}\label{thm-upper} Assuming 
	\begin{align}
	&\rho(\tilde{Q}_1,\tilde{Q}_2,Q_0)\leq \min\big\{\chi_2(\tilde{Q}_1\|Q_0),\chi_2(\tilde{Q}_2\|Q_0)\big\},\label{conv-cons1}
	\end{align}
	 $L_{\TV}^*(0^+,\delta)$ is upper bounded as follows:
	\begin{IEEEeqnarray}{rCl}
		L_{\TV}^*(0^+,\delta) \leq   \frac{2\Phi^{-1}\left(\frac{1+\delta}{2}\right)}{\sqrt{\Delta}}\cdot \max_{s\in\{1,2\}} D(\tilde{Q}_{s}\|Q_0).\label{L-thm2}
	\end{IEEEeqnarray}
\end{theorem}

The above upper bound is tight in the following special case.

\begin{theorem}[Optimality Result]\label{opt-thm} Assuming \eqref{conv-cons1}
	and
	\begin{align}
	D(\tilde{Q}_1\|Q_0)&=D(\tilde{Q}_2\|Q_0)\triangleq \mathbb{D}.\label{op-cons2}
	\end{align}
	Then,
	\begin{align}
	L_{\TV}^*(0^+,\delta)=\frac{2\Phi^{-1}\left(\frac{1+\delta}{2}\right)}{\sqrt{\Delta}}\cdot \mathbb{D}.\label{L_optimal}
	\end{align}
\end{theorem}
\begin{IEEEproof} The achievability follows from Corollary~\ref{coro} and considering the fact that assumptions in \eqref{conv-cons1} and \eqref{op-cons2} imply the first clause of $\mathsf{L}$ in \eqref{L-def1} holds. The converse follows from Theorem~\ref{thm-upper} and using the assumption~\eqref{op-cons2}. 
\end{IEEEproof}

\begin{IEEEproof}[Proof of Theorem~\ref{thm-upper}] The proof is sketched as follows. First, we relate the total variation distance to false alarm and missed detection probabilities of a hypothesis testing problem. Then, we upper bound these probabilities which  gives a lower bound on the total variation distance. Finally, the proof is concluded by showing that there exists at least one codebook which satisfies the desired property on the total variation distance and also by applying Fano's inequality.

\underline{\textit{Preparations}}: For any two codebooks 
\begin{align}\mathcal{C}_s=\left\{x_s^n(m,k)\colon m\in\mathcal{M},k\in\mathcal{K}\right\},\hspace{1cm} s\in\{1,2\},\end{align}
we define the following parameters:
\begin{IEEEeqnarray}{rCl}
	\mu_{\Lo,n}&\triangleq&\min_{s\in\{1,2\}}\;\displaystyle\min_{\substack{m\in\mathcal{M}}}\;\displaystyle\min_{\substack{k\in\mathcal{K}}}\; \frac{1}{n}\w\left(x_s^n(m,k)\right),\\
	\mu_{\Hi,n}&\triangleq&\max_{s\in\{1,2\}}\;\displaystyle\max_{\substack{m\in\mathcal{M}}}\;\displaystyle\max_{\substack{k\in\mathcal{K}}}\;\frac{1}{n}\w\left(x_s^n(m,k)\right),\label{mu-def}
\end{IEEEeqnarray}
and 
\begin{IEEEeqnarray}{rCl}
	\omega_{\Lo,n}\triangleq n\mu_{\Lo,n},\qquad \omega_{\Hi,n}\triangleq n\mu_{\Hi,n}.
	\end{IEEEeqnarray}
Let
\begin{subequations}\label{def-pri}
	\begin{IEEEeqnarray}{rCl}
		\D_{s}&\triangleq&\mathbb{E}_{Q_0}\left[ \frac{(\tilde{Q}_{s}(Y)-Q_0(Y)) (\tilde{Q}_{1}(Y)-\tilde{Q}_{2}(Y) )}{Q_{0}(Y)^2}\right],\\
			\Gamma_s &\triangleq &\mathbb{E}_{Q_0}\left[ \frac{(\tilde{Q}_{s}(Y)-Q_0(Y))(\tilde{Q}_{1}(Y)-\tilde{Q}_{2}(Y) )^2}{Q_0(Y)^3}\right],\\
		\V_{s}^* &\triangleq&  \Delta+\mu_{\Hi,n}\cdot |\Gamma_s|,
	\end{IEEEeqnarray}
\end{subequations}
and define
\begin{IEEEeqnarray}{rCl}
	\tau \triangleq \frac{n\mu_{\Lo,n}}{2}\left(\D_{2}+\D_{1}\right).\label{step15}
\end{IEEEeqnarray}

\underline{\textit{Total Variation Distance and a Hypothesis Testing Problem}}: We relate the total variation distance to a hypothesis testing problem as follows. Consider the following inequality:
\begin{IEEEeqnarray}{rCl}
	d_{\TV}(Q_1^n,Q_2^n)\geq 1-\alpha_n-\beta_n,\label{dtv-ineqaulity}
\end{IEEEeqnarray}
where $\alpha_n$ and $\beta_n$ are false alarm and missed detection probabilities of a (possibly suboptimal) hypothesis test for the following setup. Under the null hypothesis $\mathcal{H}=1$, $Y^n$ is distributed according to $Q_1^n$ and under the alternative hypothesis $\mathcal{H}=2$, it is distributed according to $Q_2^n $. Define the following suboptimal test:
\begin{align}
\mathcal{A}(y^n)\triangleq\mathbbm{1}\left\{\sum_{i=1}^n\mathcal{T}_{\test}(y_i)> \tau\right\},
\end{align}
where
\begin{align}
\mathcal{T}_{\text{\test}}(y_i)\triangleq \frac{\tilde{Q}_{1}(y_i)-\tilde{Q}_{2}(y_i)}{Q_{0}(y_i)}.\label{step44}
\end{align}
The following lemma provides upper bounds on the false alarm and missed detection probabilities of this hypothesis test.
\begin{lemma}\label{lem2} We have:
	\begin{align}
	\alpha_n \! \le \!  1-\Phi\left(\frac{1}{2}\mu_{\Lo,n}\sqrt{n\Delta}  \right) \! + \! \frac{\sqrt{n}\mu_{\Lo,n}\mu_{\Hi,n}|\Gamma_1|}{4\sqrt{2\pi\Delta}} \! + \!  O\Big(\frac{1}{\sqrt{n}}\Big).\label{step160}
	\end{align}
	and 
	\begin{align}
	\beta_n \! \le \!  1-\Phi\left(\frac{1}{2}\mu_{\Lo,n}\sqrt{n\Delta}  \right) \! + \! \frac{\sqrt{n}\mu_{\Lo,n}\mu_{\Hi,n}|\Gamma_2|}{4\sqrt{2\pi\Delta}}  \! + \!  O\Big(\frac{1}{\sqrt{n}}\Big).\label{step170}
	\end{align}
\end{lemma}
\begin{IEEEproof} See Appendix~\ref{lem2-proof}.
\end{IEEEproof}

Combining \eqref{step160} and \eqref{step170} with \eqref{dtv-ineqaulity}, we get:

\begin{IEEEeqnarray}{rCl}
	&&d_{\TV}\left(Q_1^n,Q_2^n\right)\geq 2\Phi\left(\frac{1}{2}\mu_{\Lo,n}\sqrt{n\Delta} \right)-1-\frac{\sqrt{n}\mu_{\Lo,n}\mu_{\Hi,n}(|\Gamma_1|+|\Gamma_2|)}{4\sqrt{2\pi\Delta}}+O\left(\frac{1}{\sqrt{n}}\right),\label{step42b}
\end{IEEEeqnarray}
which can be equivalently written as follows:
\begin{IEEEeqnarray}{rCl}
	&&d_{\TV}\left(Q_1^n,Q_2^n\right)\geq 2\Phi\left(\frac{1}{2}\omega_{\Lo,n}\sqrt{\frac{\Delta}{n}}  \right)-1-\frac{\omega_{\Lo,n}\omega_{\Hi,n}(|\Gamma_1|+|\Gamma_2|)}{4n\sqrt{2n\pi\Delta}}+O\left(\frac{1}{\sqrt{n}}\right).\label{step42}
\end{IEEEeqnarray}
The proof is continued by showing that there exists a sub-codebook which satisfies \eqref{step42} and its size is at least $\frac{|\mathcal{M}|\cdot |\mathcal{K}|}{\sqrt{n}}$ and it is concluded by applying Fano's inequality. See Appendix~\ref{existence-proof} for details.
\end{IEEEproof}

\begin{remark} Assume that \eqref{conv-cons1} and \eqref{op-cons2} hold. Then, the choice of $\gamma_1=\gamma_2$ minimizes the RHS of \eqref{key-ineq1} and maximizes the RHS of \eqref{key-ineq2} in Theorem~\ref{key-thm}. Thus, it can be seen a small key-length suffices for the optimal setup of Theorem~\ref{opt-thm}.
\end{remark}

\subsection{Examples}

In this section, we provide examples that satisfy both conditions of Theorem~\ref{opt-thm} in \eqref{conv-cons1} and \eqref{op-cons2} and discuss scenarios in which we can establish optimal covert communication rates.

\begin{figure}[t]
	\centering
	\includegraphics[scale=0.25]{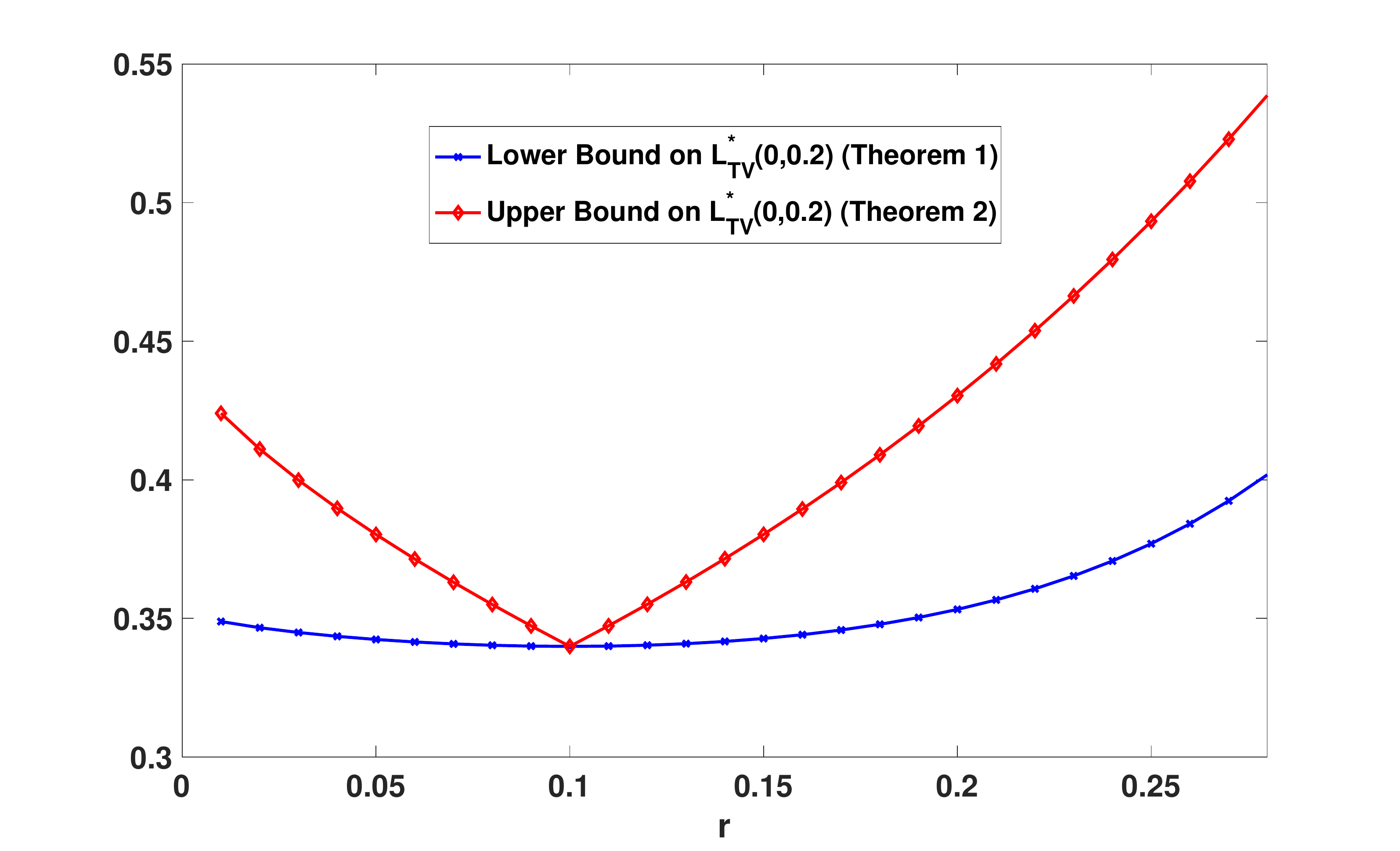}
	\caption{Lower and upper bounds on $L_{\TV}^*(0,0.2)$.}
	\label{figure3}
\end{figure}

First, consider the setup of the example in Fig.~\ref{figure2}. This example satisfies the constraints~\eqref{conv-cons1} and~\eqref{op-cons2}. Choosing $\delta=0.2$ for this example, Theorem~\ref{opt-thm} yields $L_{\TV}^*(0,0.2)=0.3399$. For other values of $r\triangleq\tilde{Q}_2(0)=\tilde{Q}_2(1)$, the upper and lower bounds on $L_{\TV}^*(0,0.2)$ are plotted in Fig.~\ref{figure3}.  It can also be verified from the figure that the two bounds match for $r=0.1$.\\

Next, consider the following Gaussian setup. Let $\tilde{Q}_s$ and $Q_0$ be $\mathcal{N}(1,\sigma_s^2)$ and $\mathcal{N}(0,\sigma_0^2)$, respectively, where $\sigma_s^2,\sigma_0^2\geq 0$ and $\frac{4}{3}\sigma_0^2\geq\sigma_1^2\geq \sigma_2^2$. Define the following parameters: 
\begin{IEEEeqnarray}{rCl}
	D_{\text{G}} &\triangleq & D(\tilde{Q}_s\|Q_0)=\frac{1}{2}\log \frac{\sigma_0^2}{\sigma_s^2}+\frac{\log e}{2\sigma_0^2}\cdot (\sigma_s^2+1),\qquad s\in\{1,2\},\\
	\chi_{\text{G},s} &\triangleq& \chi_2\left(\tilde{Q}_s\|Q_0\right)= \frac{\sigma_0}{\sqrt{2}\sigma_s}\;\exp\left(\frac{\frac{1}{2\sigma_0^2}}{1-\frac{\sigma_s^2}{2\sigma_0^2}}\right)\frac{1}{\sqrt{1-\frac{\sigma_s^2}{2\sigma_0^2}}}-1,\\
	\rho_{\text{G}} &\triangleq & \rho(\tilde{Q}_1,\tilde{Q}_2,Q_0) = 	\frac{\sigma_0}{\sqrt{\sigma_1^2+\sigma_2^2}}\exp\left( \frac{ \frac{1}{2\sigma_0^2}}{1-\frac{\sigma_1^2\sigma_2^2}{\sigma_0^2(\sigma_1^2+\sigma_2^2)}} \right)\cdot\frac{1}{\sqrt{1-\frac{\sigma_1^2\sigma_2^2}{\sigma_0^2(\sigma_1^2+\sigma_2^2)}}}-1,\\
	\Delta_{\text{G}}&\triangleq &\chi_{\text{G},1}+\chi_{\text{G},2}-2\rho_{\text{G}}.
\end{IEEEeqnarray}
 For this Gaussian setup, the conditions \eqref{conv-cons1} and \eqref{op-cons2} of Theorem~\ref{opt-thm} translate to the following: 	
\begin{IEEEeqnarray}{rCl}
	\rho_{\text{G}}\leq \min\{\chi_{\text{G},1},\chi_{\text{G},2}\}.\label{Gaus-cons2}
\end{IEEEeqnarray}
and
\begin{IEEEeqnarray}{rCl}
\sigma_0^2= \frac{\log e\cdot (\sigma_2^2-\sigma_1^2)}{	\log \frac{\sigma_2^2}{\sigma_1^2}} ,\label{Gaus-cons}
\end{IEEEeqnarray}
$\\$
Under the above assumptions, we specialize Theorem~\ref{opt-thm} to the Gaussian setup. 
\begin{corollary}\label{Gaus-cor} For any nonnegative $\sigma_s^2$ and $\sigma_0^2$ such that \begin{align}\frac{4}{3}\sigma_0^2\geq\sigma_1^2\geq \sigma_2^2,\label{cond3}\end{align} and conditions \eqref{Gaus-cons2}--\eqref{Gaus-cons} are satisfied, we have
\begin{align}
L_{\TV}^*(0^+,\delta)=\frac{2\Phi^{-1}\left(\frac{1+\delta}{2}\right)}{\sqrt{\Delta_{\text{G}}}}\cdot D_{\text{G}}.\label{L_optimal-Gaus}
\end{align}
\end{corollary}
\begin{IEEEproof} See Appendix~\ref{Gaus-proof}.
	\end{IEEEproof}

\begin{figure}[t]
	\centering
	\includegraphics[scale=0.3]{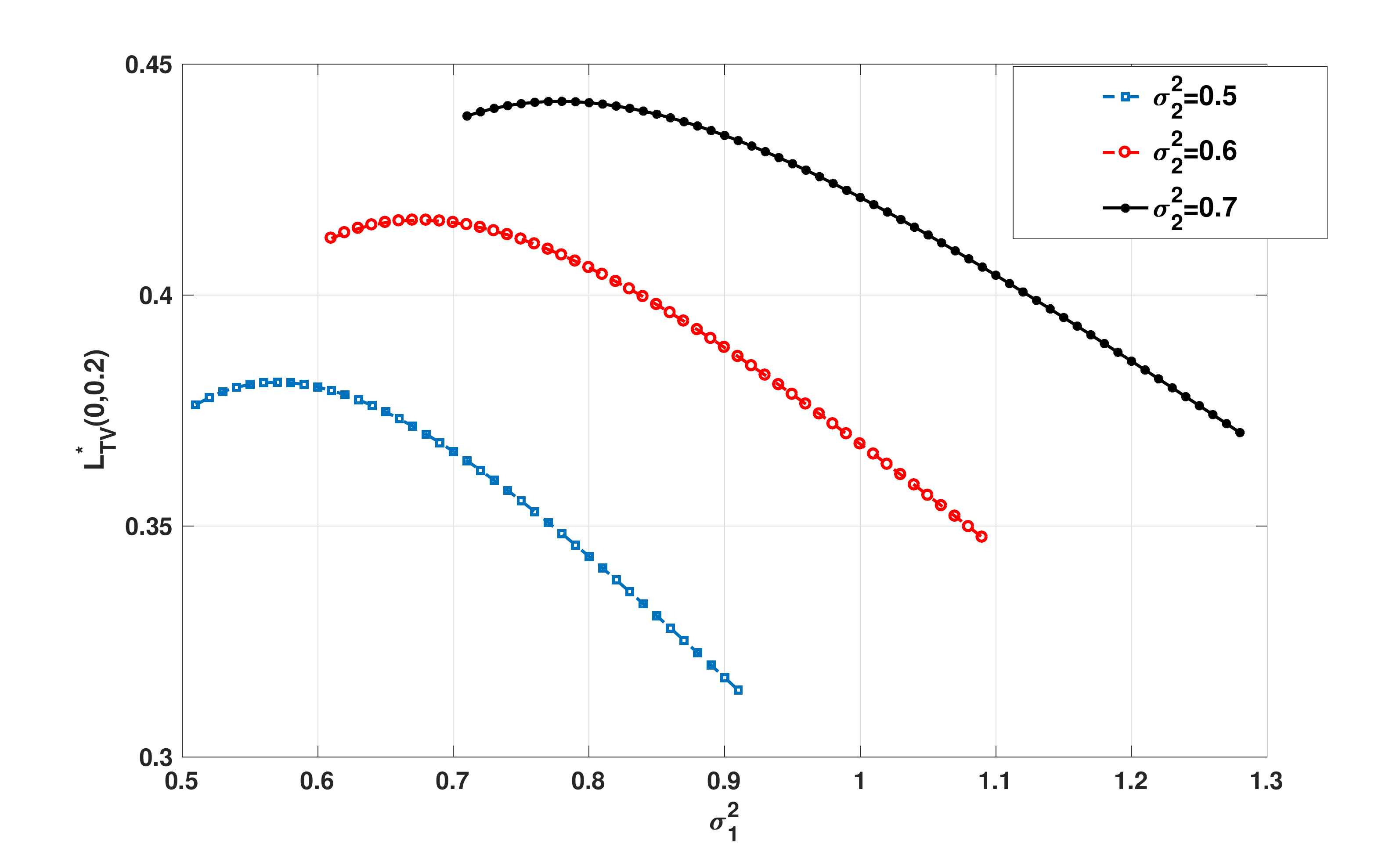}
	\caption{$L_{\TV}^*(0^+,0.2)$ for the Gaussian example.}
	\label{figure4}
\end{figure}

For the proposed Gaussian setup, Fig.~\ref{figure4} shows the optimal $L_{\TV}^*(0^+,0.2)$ versus $\sigma_1^2$ for different values of $\sigma_2^2$. The variance $\sigma_1^2$ (resp.\ $\sigma_2^2$) corresponds to the channel state with a larger (resp.\ smaller) variance of noise. Notice that for each value of $\sigma_2^2$, there exists a non-empty  interval of $\sigma_1^2$ that simultaneously satisfies the three conditions \eqref{Gaus-cons2}--\eqref{cond3}.   For $\sigma_2^2=0.5$ and $\sigma_1^2>0.57$, $L_{\TV}^*(0,0.2)$ is a decreasing function of $\sigma_1^2$. For other values of $\sigma_2^2$, a similar observation holds. The optimal covert communication rate increases as $\sigma_2^2$ increases.

\subsection{Concealing the message transmission vs concealing the channel state}\label{subsec:concealing}
To conclude this section, we revisit the discussion concerning the difference of concealing the message transmission versus concealing the channel state.  In this discussion, we remove the assumption stated in Remark~\ref{rem2} and for simplicity, we assume that the length of the key is infinite. 

 Recall the example of Fig.~\ref{figure5}. The following observations are in order:
 \begin{enumerate}
 \item As mentioned before, the assumption of Remark~\ref{rem2} does not hold, because $W_1(2|1)\neq 0$, while $W_1(2|0)=0$ and also $W_2(2|1)\neq 0$, while $W_2(2|0)=0$. Thus our results from the previous subsections, do not hold as the assumptions are violated.
 \item
  Consider the achievable scheme of the previous section with the choice $\mu_{1,n}=\mu_{2,n}\triangleq \mu_{n}$. As shown in Appendix~\ref{new-ex-proof}, under the total variation covertness criterion, if the number of messages $|\mathcal{M}|$ satisfies
\begin{align}
\log |\mathcal{M}| \leq \frac{\sqrt{\ln \frac{1}{1-\delta}}}{2}\cdot \sqrt{n}\log n+o\left(\sqrt{n}\log n\right),\label{step3000}
\end{align}
then the desired constraints on the maximum error probability and covertness metric are satisfied.
The above result shows that  it is possible to  communicate $\Omega(\sqrt{n}\log n)$ bits over $n$ channel uses covertly in the sense of concealing of the channel state. However, covert communication in the sense of concealing message transmission is not feasible for this example in the sense that the KL-divergence $D\left(Q_s^n\|Q_0^{\times n}\right)$ tends to infinity as $n\to \infty$.
 \item   As mentioned in the previous item, the communication rate is at least of the order of $\Omega\left(\frac{\log n}{\sqrt{n}}\right)$, which is somewhat surprising,  because the communication rate is usually of order $\Theta\left(\frac{1}{\sqrt{n}}\right)$ for covert communication problems~\cite{Bloch,Ligong}, as we have observed in our results in the previous subsections. This again shows the importance of the absolutely continuous assumption in Remark \ref{rem2}.
 \item Although we will not discuss the details, it is worth  mentioning that unlike the achievable scheme of the previous subsection, the choice $\mu_{1,n}=\mu_{2,n}$ is the unique choice which enables sending a positive number of bits under the covert communication constraint.
 \end{enumerate}

\section{Conclusion}\label{sec:conclusion}

In this paper, we considered covert communication over a compound channel. In this setting, the covertness is defined as the ability of a malicious party in distinguishing the channel state. We provided inner and outer bounds to the optimal throughput-key region and showed that the square-root law holds for this setup. For a special case, the bounds match and establish the optimal throughput. Some examples, including a Gaussian setup, are given to discuss the bounds. 

A promising avenue for future work is to study covert communication over a channel with an i.i.d.\ state $S^n$ available at  the encoder, the decoder or both. The distribution that generates $S^n$ can either be $P_{S_1}$ or $P_{S_2}$. The main goal in this stochastic--as opposed to deterministic in this paper---setting is to design a code to remain covert such that an adversary who is observing the channel output does not learn the channel state distribution, i.e.,  $P_{S_1}$ or $P_{S_2}$. The problem of covert communication over a channel with state has been previously studied in \cite{Lee}. One can consider an alternative formulation to conceal the distribution of state instead of the presence or absence of message transmission as was done in \cite{Lee}.

\appendices

\section{Proof of Lemma \ref{lem-ach}}\label{lem-ach-proof}

First, we present the following theorem (Berry-Esseen Theorem) which will be used repeatedly in the proofs, later. 
\begin{theorem}[Thm~1.6 in \cite{Vincent-book}]\label{BE-thm} Let $X^n\triangleq (X_1,\ldots,X_n)$ be a collection of $n$ independent random variables where each random variable has a zero mean, variance $\sigma_i^2\triangleq \mathbb{E}\left[X_i^2\right]>0$ and third absolute moment $T_i\triangleq \mathbb{E}\left[|X_i|^3\right]<\infty$. Define the average variance and average third absolute moment as $\sigma^2\triangleq \frac{1}{n}\sum_{i=1}^n\sigma_i^2$ and $T\triangleq \frac{1}{n}\sum_{i=1}^nT_i$, respectively. Then, we have:
	\begin{IEEEeqnarray}{rCl}
		\sup_{a\in\mathbb{R}}\Bigg|\mathbb{P}\left(\frac{1}{\sigma\sqrt{n}}\sum_{i=1}^nX_i<a \right)-\Phi(a)\Bigg|\leq \frac{6T}{\sigma^3\sqrt{n}}.
	\end{IEEEeqnarray}
\end{theorem}
Now, consider the following identity for the total variation distance:
\begin{IEEEeqnarray}{rCl}
	\hspace{-0.5cm}	d_{\TV}\left(Q_1^{\times n},Q_2^{\times n}\right)&= &\mathbb{P}_{Q_1^{\times n}}\left(\log\left(\frac{Q_1^{\times n}}{Q_2^{\times n}}\right)\geq 0\right)-\mathbb{P}_{Q_2^{\times n}}\left(\log\left(\frac{Q_1^{\times n}}{Q_2^{\times n}}\right)\geq 0\right)\\[2ex]
	&= &\mathbb{P}_{Q_1^{\times n}}\left(\sum_{i=1}^n\log\frac{Q_{1,n}(Y_i)}{Q_{2,n}(Y_i)}\geq 0\right)-\mathbb{P}_{Q_2^{\times n}}\left(\sum_{i=1}^n\log\frac{Q_{1,n}(Y_i)}{Q_{2,n}(Y_i)}\geq 0\right).
	\label{step18}
\end{IEEEeqnarray}
Now, we bound each of the probabilities in \eqref{step18}. First, we analyze the first probability. The mean and variance of  $\sum_{i=1}^n\log\frac{Q_{1,n}(Y_i)}{Q_{2,n}(Y_i)}$ are given by the following:
\begin{IEEEeqnarray}{rCl}
	&&\mathbb{E}\left[\sum_{i=1}^n\log\frac{Q_{1,n}(Y_i)}{Q_{2,n}(Y_i)}\right]=nD(Q_{1,n} \| Q_{2,n})\triangleq n\text{d}_1,\end{IEEEeqnarray}
and
\begin{IEEEeqnarray}{rCl}
	\mathbb{V}\left[\sum_{i=1}^n\log\frac{Q_{1,n}(Y_i)}{Q_{2,n}(Y_i)}\right]&=&n \left(\sum_yQ_{1,n}(y)\left(\log\frac{Q_{1,n}(y)}{Q_{2,n}(y)}\right)^2-D^2(Q_{1,n} \| Q_{2,n})\right)\\
	&\triangleq & n\text{v}_1.\label{step38}
\end{IEEEeqnarray}
Employing  Theorem~\ref{BE-thm}, we obtain
\begin{IEEEeqnarray}{rCl}
	\mathbb{P}_{Q_1^{\times n}}\left(\sum_{i=1}^n\log\frac{Q_{1,n}(Y_i)}{Q_{2,n}(Y_i)}\geq 0\right) \leq \Phi\left(\frac{\sqrt{n}\text{d}_1}{\sqrt{\text{v}_1}}\right)+\frac{6\text{t}_1}{\sqrt{n}\text{v}_1^{\frac{3}{2}}},\label{step19}
\end{IEEEeqnarray}
where
\begin{align}
\text{t}_1\triangleq \mathbb{E}_{Q_{1,n}}\left[\Bigg|\log \frac{Q_{1,n}(Y)}{Q_{2,n}(Y)}-D(Q_{1,n}\|Q_{2,n})\Bigg|^3\right].
\end{align}
Similarly, we get:
\begin{IEEEeqnarray}{rCl}
	\mathbb{P}_{Q_2^{\times n}}\left(\sum_{i=1}^n\log\frac{Q_{1,n}(Y_i)}{Q_{2,n}(Y_i)}\geq 0\right) \geq 1-\Phi\left(\frac{\sqrt{n}\text{d}_2}{\sqrt{\text{v}_2}}\right)-\frac{6\text{t}_2}{\sqrt{n}\text{v}_2^{\frac{3}{2}}},\nonumber\\
	\label{step20}
\end{IEEEeqnarray}
where we define:
\begin{IEEEeqnarray}{rCl}
	\text{d}_2&\triangleq & D(Q_{2,n}\|Q_{1,n}),\\
	\text{v}_2 &\triangleq & \sum_yQ_{2,n}(y)\left(\log\frac{Q_{1,n}(y)}{Q_{2,n}(y)}\right)^2-D^2(Q_{2,n} \| Q_{1,n}),\\
	\text{t}_2&\triangleq &\mathbb{E}_{Q_{2,n}}\left[\Bigg|\log \frac{Q_{1,n}(Y)}{Q_{2,n}(Y)}+D(Q_{2,n}\|Q_{1,n})\Bigg|^3\right].
\end{IEEEeqnarray}
Now, consider the following set of approximations:
\begin{IEEEeqnarray}{rCl}
	&&\hspace{-0.3cm}\text{d}_1 = D(Q_{1,n}\|Q_{2,n})\\
	&=&\sum_y Q_{1,n}(y)\log \frac{Q_{1,n}(y)}{Q_{2,n}(y)}\\
	&=&\mathbb{E}_{Q_{2,n}}\left[\frac{Q_{1,n}}{Q_{2,n}}\log \frac{Q_{1,n}}{Q_{2,n}}\right]\\
	&=&\mathbb{E}_{Q_{2,n}}\left[\frac{(1-\mu_{1,n})Q_0+\mu_{1,n}\bar{Q}_1}{(1-\mu_{2,n})Q_0+\mu_{2,n}\bar{Q}_2}\log \frac{(1-\mu_{1,n})Q_0+\mu_{1,n}\bar{Q}_1}{(1-\mu_{2,n})Q_0+\mu_{2,n}\bar{Q}_2}\right]\\
	&=&\mathbb{E}_{Q_{2,n}}\left[\left(1+\frac{\mu_{1,n}\big(\bar{Q}_1-Q_0\big)-\mu_{2,n} \big(\bar{Q}_{2}-Q_0\big)}{(1-\mu_{2,n})Q_0+\mu_{2,n}\bar{Q}_2}\right)\log \left(1+\frac{\mu_{1,n}\big(\bar{Q}_1-Q_0\big)-\mu_{2,n} \big(\bar{Q}_{2}-Q_0\big)}{(1-\mu_{2,n})Q_0+\mu_{2,n}\bar{Q}_2}\right)\right]\label{step24}\\
		&=&\mathbb{E}_{Q_{2,n}}\Bigg[\left(1+\frac{\mu_{1,n}\big(\bar{Q}_1-Q_0\big)-\mu_{2,n} \big(\bar{Q}_{2}-Q_0\big)}{(1-\mu_{2,n})Q_0+\mu_{2,n}\bar{Q}_2}\right)\times \nonumber\\&&\hspace{2cm} \left(\frac{\mu_{1,n}\big(\bar{Q}_1-Q_0\big)-\mu_{2,n} \big(\bar{Q}_{2}-Q_0\big)}{(1-\mu_{2,n})Q_0+\mu_{2,n}\bar{Q}_2}-\frac{1}{2}\left(\frac{\mu_{1,n}\big(\bar{Q}_1-Q_0\big)-\mu_{2,n} \big(\bar{Q}_{2}-Q_0\big)}{(1-\mu_{2,n})Q_0+\mu_{2,n}\bar{Q}_2}\right)^2\right)\Bigg]+o\left(\frac{1}{n}\right)\nonumber\\\\
	&=&\frac{1}{2} \mathbb{E}_{Q_{2,n}}\left[\left(\frac{\mu_{1,n}\big(\bar{Q}_1-Q_0\big)-\mu_{2,n} \big(\bar{Q}_{2}-Q_0\big)}{(1-\mu_{2,n})Q_0+\mu_{2,n}\bar{Q}_{2}}\right)^2\right]+o\left(\frac{1}{n}\right)\\[2ex]
	&=&\frac{1}{2} \mathbb{E}_{Q_0}\left[\left(\frac{\mu_{1,n}\big(\bar{Q}_1-Q_0\big)-\mu_{2,n} \big(\bar{Q}_{2}-Q_0\big)}{Q_0}\right)^2\right]+o\left(\frac{1}{n}\right)\label{step25b}\\[1ex]
	&=&\frac{1}{2}\mu_{1,n}^2\chi_2\left(\bar{Q}_1\|Q_0\right)+\frac{1}{2}\mu_{2,n}^2\chi_2\left(\bar{Q}_2\|Q_0\right)-\mu_{1,n}\mu_{2,n}\rho\left(\bar{Q}_1,\bar{Q}_2,Q_0\right)+o\left(\frac{1}{n}\right),\label{step21}
\end{IEEEeqnarray}
where \eqref{step24} follows from the Taylor expansion $\log(1+x)=x-\frac{1}{2}x^2+o(x^2)$ at $x=0$; \eqref{step25b} follows because \begin{align}Q_{2,n}=(1-\mu_{2,n})Q_0+\mu_{2,n}\bar{Q}_{2}\to Q_0.\end{align}
Now, define:
\begin{IEEEeqnarray}{rCl}
	\bar{\Omega}\left(\mu_{1,n},\mu_{2,n}\right)&\triangleq& \mu_{1,n}^2\chi_2\left(\bar{Q}_1\|Q_0\right)+\mu_{2,n}^2\chi_2\left(\bar{Q}_2\|Q_0\right)-2\mu_{1,n}\mu_{2,n}\rho\left(\bar{Q}_1,\bar{Q}_2,Q_0\right).
\end{IEEEeqnarray}
The equality \eqref{step21} can be expressed in terms of $\bar{\Omega}\left(\mu_{1,n},\mu_{2,n}\right)$ as the following:
\begin{IEEEeqnarray}{rCl}
	\text{d}_1=\frac{1}{2}\bar{\Omega}\left(\mu_{1,n},\mu_{2,n}\right)+o\left(\frac{1}{n}\right).
\end{IEEEeqnarray}
Next, we consider the following set of approximations:
\begin{IEEEeqnarray}{rCl}
	\sum_yQ_1(y)\left(\log\frac{Q_1(y)}{Q_2(y)}\right)^2
	&=&\sum_y \left((1-\mu_{1,n})Q_0(y)+\mu_{1,n}\bar{Q}_{1}(y)\right)\left(\log \frac{(1-\mu_{1,n})Q_0(y)+\mu_{1,n}\bar{Q}_{1}(y)}{(1-\mu_{2,n})Q_0(y)+\mu_{2,n}\bar{Q}_{2}(y)}\right)^2\\[1.5ex]
	&=&\sum_y \left((1-\mu_{1,n})Q_0(y)+\mu_{1,n}\bar{Q}_{1}(y)\right)\left(\log \left(1+\frac{\mu_{1,n}\big(\bar{Q}_1-Q_0 \big)-\mu_{2,n} \big(\bar{Q}_{2}-Q_0\big)}{(1-\mu_{2,n})Q_0(y)+\mu_{2,n}\bar{Q}_{2}(y)}\right)\right)^2\\[1.5ex]
	&=&\sum_y \left((1-\mu_{1,n})Q_0(y)+\mu_{1,n}\bar{Q}_{1}(y)\right)\left(\frac{\mu_{1,n}\big(\bar{Q}_1-Q_0 \big)-\mu_{2,n} \big(\bar{Q}_{2}-Q_0\big)}{(1-\mu_{2,n})Q_0(y)+\mu_{2,n}\bar{Q}_{2}(y)}\right)^2+o\left(\frac{1}{n}\right)\nonumber\\\label{step37b}\\[1.5ex]
	&=&\sum_y Q_0(y)\left(\frac{\mu_{1,n}\big(\bar{Q}_1-Q_0 \big)-\mu_{2,n} \big(\bar{Q}_{2}-Q_0\big)}{Q_0(y)}\right)^2+o\left(\frac{1}{n}\right)\label{step37}\\[1.5ex]
	&=& \mathbb{E}_{Q_0}\left[\left(\frac{\mu_{1,n}\big(\bar{Q}_1-Q_0 \big)-\mu_{2,n} \big(\bar{Q}_{2}-Q_0\big)}{Q_0}\right)^2\right]+o\left(\frac{1}{n}\right)\\[1.5ex]
	&=&\mu_{1,n}^2\chi_2\left(\bar{Q}_1\|Q_0\right)+\mu_{2,n}^2\chi_2\left(\bar{Q}_2\|Q_0\right)-2\mu_{1,n}\mu_{2,n}\rho\left(\bar{Q}_1,\bar{Q}_2,Q_0\right)+o\left(\frac{1}{n}\right)\\[1.5ex]
	&=&\bar{\Omega}\left(\mu_{1,n},\mu_{2,n}\right)+o\left(\frac{1}{n}\right),\label{step22}
\end{IEEEeqnarray}
where \eqref{step37b} follows from Taylor expansion $\log(1+x)=x+o(x)$ at $x=0$, \eqref{step37} follows because $$(1-\mu_{s,n})Q_0(y)+\mu_{s,n}\bar{Q}_{s}(y)\to Q_0(y),$$
as $\mu_{s,n}\to 0$.
Combining \eqref{step19}, \eqref{step21}, \eqref{step22} and definition of $\text{v}_1$ in \eqref{step38}, we have: 
\begin{align}
\Phi\left(\frac{\sqrt{n}\text{d}_1}{\sqrt{\text{v}_1}}\right)&= \Phi\left(\frac{1}{2}\sqrt{n\bar{\Omega}\left(\mu_{1,n},\mu_{2,n}\right)}\right)+o\left(\frac{1}{n}\right).\label{step39}
\end{align}
Similarly, 
\begin{align}
\Phi\left(\frac{\sqrt{n}\text{d}_2}{\sqrt{\text{v}_2}}\right)&= \Phi\left(\frac{1}{2}\sqrt{n\bar{\Omega}\left(\mu_{1,n},\mu_{2,n}\right)}\right)+o\left(\frac{1}{n}\right).\label{step40}
\end{align}
Following similar steps leading to \eqref{step22}, one can show that $\text{t}_s=O\left(\frac{1}{n^{\frac{3}{2}}}\right)$.
Combining \eqref{step18}, \eqref{step19}, \eqref{step20}, \eqref{step39} and \eqref{step40}, and considering the fact that $\text{v}_s=\Theta\left(\frac{1}{n}\right)$, we obtain the following:
\begin{IEEEeqnarray}{rCl}
	d_{\TV}\left(Q_1^{\times n},Q_2^{\times n}\right) &\leq& \Phi\left(\frac{\sqrt{n}\text{d}_1}{\sqrt{\text{v}_1}}\right)+\Phi\left(\frac{\sqrt{n}\text{d}_2}{\sqrt{\text{v}_2}}\right)-1+\frac{6\text{t}_1}{\sqrt{n}\text{v}_1^{\frac{3}{2}}}+\frac{6\text{t}_2}{\sqrt{n}\text{v}_2^{\frac{3}{2}}} \label{eqn:81}
	\\&\leq& 2\Phi\left(\frac{1}{2}\sqrt{n\bar{\Omega}\left(\mu_{1,n},\mu_{2,n}\right)}\right)-1+O\left(\frac{1}{\sqrt{n}}\right).\label{step25}
\end{IEEEeqnarray}
Considering the fact that $\gamma_s=\sqrt{n}\mu_{s,n}$, completes the proof of the lemma.

\section{Proof of Lemma~\ref{lem-key}}\label{lem-key-proof}

First, we state the following lemma.
\begin{lemma} [Corollary VII.2 in \cite{Cuff}] For all positive $\eta_s$, we have:
	\begin{IEEEeqnarray}{rCl}
		\mathbb{E}_{\mathcal{C}_s}\left[d_{\TV}\left(Q_s^n,Q_s^{\times n}\right)\right]\leq \frac{1}{2}\sqrt{\frac{2^{\eta_s}}{|\mathcal{M}|\cdot |\mathcal{K}|}}+\frac{1}{2}\mathbb{P}_{W_s^nP_s^{\times n}}\left( \sum_{i=1}^n\log \frac{W_s(Y_i|X_i)}{Q_{s,n}(Y_i)} \geq \eta_s \right).\label{lem-exp}
	\end{IEEEeqnarray}
\end{lemma}
Choose 
\begin{IEEEeqnarray}{rCl}
	\eta_s = n\left(1+\frac{\kappa}{2}\right)I(P_{s,n},W_s),
	\end{IEEEeqnarray}
for some positive $\kappa$.
Notice that from \cite{Ligong}, the following approximations hold:
\begin{IEEEeqnarray}{rCl}
	I(P_s,W_s)&=&\frac{1}{\sqrt{n}}\gamma_s
	D(W_s \| Q_0|\bar{P}_s)
	+o\left(\frac{1}{\sqrt{n}}\right)\\
	&\triangleq &\frac{\gamma_s}{\sqrt{n}}\bar{\mathbb{D}}_s+o\left(\frac{1}{\sqrt{n}}\right),\label{step1000}
	\end{IEEEeqnarray}
Thus, 
\begin{IEEEeqnarray}{rCl}
	\eta_s = \gamma_s\sqrt{n}\left(1+\frac{\kappa}{2}\right)\bar{\mathbb{D}}_s+o(\sqrt{n}).
	\end{IEEEeqnarray}
Also, we have
\begin{IEEEeqnarray}{rCl}
	\sum_{x}\sum_y P_{s,n}(x)W_s(y|x)\left(\log \frac{W_s(y|x)}{Q_{s,n}(y)}\right)^2
	&=& \frac{\gamma_s}{\sqrt{n}}\sum_{x\neq 0}\sum_y \bar{P}_{s}(x)W_s(y|x)\left(\log \frac{W_s(y|x)}{Q_0(y)}\right)^2+o\left( \frac{1}{\sqrt{n}}\right)\\
	&\triangleq & \frac{\gamma_s}{\sqrt{n}}\bar{\mathbb{V}}_s+o\left( \frac{1}{\sqrt{n}}\right).\label{step1001}
	\end{IEEEeqnarray}
Now, consider the probability term in \eqref{lem-exp} as follows:

\begin{IEEEeqnarray}{rCl}
	&&\hspace{-1cm}\mathbb{P}_{W_s^nP_s^{\times n}}\left(\sum_{i=1}^n\log \frac{W_s(Y_i|X_i)}{Q_{s,n}(Y_i)} \geq \eta_s \right) \\
	&=&\mathbb{P}_{W_s^nP_s^{\times n}}\left(\sum_{i=1}^n\log \frac{W_s(Y_i|X_i)}{Q_{s,n}(Y_i)} \geq n(1+\frac{\kappa}{2})I(P_{s,n},W_s) \right)\\
	&=&\mathbb{P}_{W_s^nP_s^{\times n}}\left( \sum_{i=1}^n\log \frac{W_s(Y_i|X_i)}{Q_{s,n}(Y_i)}-nI(P_{s,n},W_s) \geq \frac{n\kappa}{2} I(P_{s,n},W_s) \right)\\
	&=&\mathbb{P}_{W_s^nP_s^{\times n}}\left(\sum_{i=1}^n\log \frac{W_s(Y_i|X_i)}{Q_{s,n}(Y_i)}-nI(P_{s,n},W_s) \geq \frac{\sqrt{n}\kappa \gamma_s\bar{\mathbb{D}}_s}{2}+o(\sqrt{n}) \right) \\
	&\leq & \mathbb{P}_{W_s^nP_s^{\times n}}\left( \Big|\sum_{i=1}^n\log \frac{W_s(Y_i|X_i)}{Q_{s,n}(Y_i)}-nI(P_{s,n},W_s)\Big| \geq \frac{\sqrt{n}\kappa \gamma_s\bar{\mathbb{D}}_s}{2}+o(\sqrt{n}) \right)\\
	&\leq & \frac{\sqrt{n}\gamma_s\bar{\mathbb{V}}_s}{4n\kappa^2\gamma_s^2\bar{\mathbb{D}}_s^2}+o\left(\frac{1}{\sqrt{n}}\right)\\&=& O\left(\frac{1}{\sqrt{n}}\right),
	\label{ineq2}
\end{IEEEeqnarray}
where the last inequality follows from Chebyshev's inequality and the fact that 
\begin{IEEEeqnarray}{rCl}
	\mathbb{V}\left[\sum_{i=1}^n\log \frac{W_s(Y_i|X_i)}{Q_{s,n}(Y_i)}-nI(P_{s,n},W_s)\right]&\leq & n\left(	\sum_{x}\sum_y P_{s,n}(x)W_s(y|x)\left(\log \frac{W_s(y|x)}{Q_{s,n}(y)}\right)^2\right)\\
	&\leq & \sqrt{n}\gamma_s\bar{\mathbb{V}}_s+o(\sqrt{n}). \label{step200}
	\end{IEEEeqnarray}
Combining \eqref{lem-exp} and \eqref{ineq2}, we get:
\begin{IEEEeqnarray}{rCl}
	\mathbb{E}_{\mathcal{C}_s}\left[d_{\TV}\left(Q_s^n,Q_s^{\times n}\right)\right] 
	&\leq &\frac{1}{2}\sqrt{\frac{2^{\eta_s}}{|\mathcal{M}|\cdot |\mathcal{K}|}}+O\left(\frac{1}{\sqrt{n}}\right).\label{total-distance}
\end{IEEEeqnarray}
Now, we choose:
\begin{IEEEeqnarray}{rCl}
	\log|\mathcal{M}|+\log|\mathcal{K}|&\geq& \gamma_s\sqrt{n}\left(1+{\kappa}\right)\bar{\mathbb{D}}_s.\label{ineq600}
\end{IEEEeqnarray}
With the above choice, we can further upper bound \eqref{total-distance} as follows:
\begin{IEEEeqnarray}{rCl}
	\mathbb{E}_{\mathcal{C}_s}\left[d_{\TV}\left(Q_s^n,Q_s^{\times n}\right)\right] 
	&= & O\left(\frac{1}{\sqrt{n}}\right).
	\label{total-distance2}
\end{IEEEeqnarray}
This completes the proof of the lemma.

\section{Proof of Lemma~\ref{PPM-thm}}\label{rate-analysis}

We provide the analysis of the rate constraint in the following. Since the ML decoder results in the smallest probability of error for a given codebook, we can bound the average probability of error using Shannon's achievability bound (threshold decoding) \cite[Thm~2]{Yury} as follows. For every $s\in\{1,2\}$ and $k\in\mathcal{K}$, the average error probability satisfies:
\begin{IEEEeqnarray}{rCl}
	\mathbb{E}\left[\bar{P}_{\mathrm{e}}(s,k)\right] \leq \mathbb{P}\left( \log \frac{W_s(Y^n|X^n)}{Q_s^{\times n}(Y^n)}\leq \log |\mathcal{M}|+\frac{1}{6}\log n \right)+n^{-\frac{1}{6}}.\label{Pe}
	\end{IEEEeqnarray}
Recall the definition of $\bar{\mathbb{V}}_s$ from \eqref{step1001} and consider the above probability:
\begin{IEEEeqnarray}{rCl}
	\mathbb{P}\left( \log \frac{W_s(Y^n|X^n)}{Q_s^{\times n}(Y^n)}\leq \log |\mathcal{M}|+\frac{1}{6}\log n \right)&\leq &	\mathbb{P}\left(\log \frac{W_s(Y^n|X^n)}{Q_s^{\times n}(Y^n)}\leq  nI(P_{s,n},W_s)-n^{\frac{1}{3}}+\frac{1}{6}\log n \right) \\
	&=&\mathbb{P}\left( \log \frac{W_s(Y^n|X^n)}{Q_s^{\times n}(Y^n)}-nI(P_{s,n},W_s)\leq  -n^{\frac{1}{3}}+\frac{1}{6}\log n \right)\\
	&\leq & \mathbb{P}\left(\Big|\log \frac{W_s(Y^n|X^n)}{Q_s^{\times n}(Y^n)}-nI(P_{s,n},W_s)\Big|\geq  n^{\frac{1}{3}}-\frac{1}{6}\log n\right) \\
	&\leq & \frac{\sqrt{n}\gamma_s\bar{\mathbb{V}}_s}{(n^{\frac{1}{3}}-\frac{1}{6}\log n)^2}\\
	&=& O\left(n^{-\frac{1}{6}}\right),
	\end{IEEEeqnarray}
where the last inequality follows from Chebyshev's inequality and the steps leading to \eqref{step200}.  This completes the proof of lemma.

\section{Proof of Existence of a Codebook With Desired Properties}\label{existence2-proof}
First, we state the following theorem which will be used later in the proof.
\begin{theorem}[McDiarmid's theorem\cite{Mc}]\label{Mc-thm} Let $\{X_k\}_{k=1}^n$ be independent random variables defined on the set $\mathcal{X}$. Consider a random variable $U=f(X^n)$ where $f$ is a function satisfying the following bounded difference property:
	\begin{IEEEeqnarray}{rCl}
		\sup_{x_1,\ldots,x_n,x'_i\in\mathcal{X}}f(x_1,\ldots,x_i,\ldots,x_n)-f(x_1,\ldots,x'_i,\ldots,x_n)\leq d_i,\qquad \forall i\in\{1,\ldots,n\},\label{eqn:bnd-diff}
	\end{IEEEeqnarray}
	for some positive numbers $d_i, 1\le i\le n$. Then, for every $r>0$, 
	\begin{IEEEeqnarray}{rCl}
		\mathbb{P}\left(U-\mathbb{E}\left[U\right]>r\right)\leq \exp\left(\frac{-2r^2}{\sum_{i=1}^n d_i^2}\right).
	\end{IEEEeqnarray}
\end{theorem}

\underline{\textit{Step 1: Probability of Satisfying the Covertness Condition}}:

Choose positive real numbers $D_n=\frac{1}{\sqrt{\log{n}}}$ and 
 $a= n^{-\frac{1}{12}}$; and define the following events:
\begin{IEEEeqnarray}{rCl}
	\mathcal{B}_s&\triangleq& \Bigg\{ \forall k\in\mathcal{K},\; \exists \mathcal{M}_{s,k}\subset \mathcal{M},\; \mathcal{M}_s=\mathcal{M}_{s,1}\times \ldots\times \mathcal{M}_{s,k}\colon  |\mathcal{M}_{s,k}|=(1-a)|\mathcal{M}|,\;\; \nonumber\\[-1ex]
	&&\hspace{2.5cm}Q_{\mathcal{M}_s}^n(y^n)\triangleq \frac{1}{|\mathcal{M}_s \| \mathcal{K}|}\;\sum_{k\in\mathcal{K}}\;\sum_{m\in\mathcal{M}_{s,k}}W_s^n(y^n|X_s^n(m,k)),\;\; d_{\TV}(Q_{\mathcal{M}_s}^n,Q_{s}^{\times n})> D_n \Bigg\},\label{event-second}
\end{IEEEeqnarray}
for $s=1,2$.

Consider the probability of this event:
\begin{IEEEeqnarray}{rCl}
	\mathbb{P}\left(\mathcal{B}_s\right) &=& \mathbb{P}\left(\forall k\in\mathcal{K}, \;\exists \mathcal{M}_{s,k}\subset \mathcal{M}\colon |\mathcal{M}_{s,k}|=(1-a)|\mathcal{M}|,\;\;d_{\TV}(Q_{\mathcal{M}_s}^n,Q_{s}^{\times n})> D_n \right)\\
	&\stackrel{(a)}{\leq} & \mathbb{P}\left(\forall k\in\mathcal{K},\;\exists \mathcal{M}_{s,k}\subset \mathcal{M}\colon |\mathcal{M}_{s,k}|=(1-a)|\mathcal{M}|,\;\;d_{\TV}(Q_{\mathcal{M}_s}^n,Q_{s}^{\times n})-\mathbb{E}\left[d_{\TV}(Q_{\mathcal{M}_s}^n,Q_{s}^{\times n})\right] >\frac{D_n}{2}\right)\\
	&\stackrel{(b)}{\leq} &
	 \sum_{\mathcal{M}_{s,k}\subset \mathcal{M}, k\in\mathcal{K}\colon\atop |\mathcal{M}_{s,k}|=(1-a)|\mathcal{M}|}\mathbb{P}\left(d_{\TV}(Q_{\mathcal{M}_s}^n,Q_{s}^{\times n})- \mathbb{E}\left[d_{\TV}(Q_{\mathcal{M}_{s}}^n,Q_{s}^{\times n})\right]>\frac{D_n}{2}\right)
	,\label{step600}
\end{IEEEeqnarray}
where 
\begin{itemize}
\item$(a)$ follows because according to \eqref{exp:dTV}, the expectation of the total variation distance is upper bounded by $O(n^{-\frac{1}{2}})\le\frac{D_n}{2}$ for sufficiently large $n$, because 
$$\log |\mathcal{M}_s|+\log |\mathcal{K}|=\log |\mathcal{M}|+\log |\mathcal{K}|+\log(1-a)\ge  (1+\kappa')\sqrt{n} \cdot \max_{s\in\{1,2\}}\gamma_s
	D(W_s \| Q_0|\bar{P}_s),$$
for some $\kappa'>0$ that approaches $\kappa$ as $n\to\infty$,
\item $(b)$ follows from the union bound.
\end{itemize}
 In order to upper bound \eqref{step600}, we use the following lemma due to Liu, {\em et. al.} \cite{jingbo} (see also \cite{Mehrdad}), showing that the total variation distance of a random codebook is concentrated around its expectation. 
 \begin{lemma}[Theorem~31 in \cite{jingbo}]\label{le:jingbo}
 Consider a channel with the probability transition $W$ and an input probability distribution $P_X$. Suppose that $P_Y$ is the output distribution of the channel with the input distribution $P_X$. Let $\mathcal{C}=\{X_1,\cdots,X_{\mathsf{M}}\}$ be a random codebook with i.i.d. codewords drawn from $P_X$ and $\hat{P}_Y(.)=\frac{1}{\mathsf{M}}\sum_{m=1}^\mathsf{M} W(.|X_m)$ be the induced output distribution by $\mathcal{C}$. Then for any $\Delta>0$, we have
 \begin{align}
 \mathbb{P}(d_{\TV}(\hat{P}_Y,P_Y)-\mathbb{E}[(d_{\TV}(\hat{P}_Y,P_Y)]>\Delta)\le \exp_e(-2\mathsf{M}\Delta^2).\label{eqn:jingbo}
 \end{align}  
 \end{lemma}

 Using the above lemma with the following specifications,
 \begin{align}
 \mathcal{C}&\leftarrow \mathcal{C}_s=\bigcup_{k\in |\mathcal{K}|}\{X^n_{s}(m,k):m\in\mathcal{M}_{s,k}\},\\
 \mathsf{M}&\leftarrow (1-a)|\mathcal{M}||\mathcal{K}|,\\
 Y&\leftarrow Y^n\\
 W&\leftarrow W^n_{s},\\
 \hat{P}_{Y}&\leftarrow Q_{\mathcal{M}_s}^n,\\
 P_Y&\leftarrow Q_{s}^{\times n},\\
 \Delta&\leftarrow \frac{D_n}{2},
 \end{align}
 we get the following upper bound on
the term \eqref{step600} as follows: 
\begin{IEEEeqnarray}{rCl}
	&&\sum_{\mathcal{M}_{s,k}\subset \mathcal{M}, k\in\mathcal{K}\colon\atop |\mathcal{M}_{s,k}|=(1-a)|\mathcal{M}|}\mathbb{P}\left(d_{\TV}(Q_{\mathcal{M}_s}^n,Q_{s}^{\times n})- \mathbb{E}\left[d_{\TV}(Q_{\mathcal{M}_{s}}^n,Q_{s}^{\times n})\right]>\frac{D_n}{2}\right) \nonumber\\&&\hspace{1cm}\leq \sum_{\mathcal{M}_{s,k}\subset \mathcal{M}, k\in\mathcal{K}\colon\atop |\mathcal{M}_{s,k}|=(1-a)|\mathcal{M}|}\exp\left(-\frac{1}{2}D_n^2(1-a)|\mathcal{M}\| \mathcal{K}|\right)\\
	&&\hspace{1cm}={|\mathcal{M}| \choose (1-a)|\mathcal{M}|}^{|\mathcal{K}|}\exp\left(-\frac{1}{2}(1-a)D_n^2|\mathcal{M} \| \mathcal{K}|\right)\\
	&&\hspace{1cm}\stackrel{(d)}{\leq} \exp\left(-|\mathcal{M} \| \mathcal{K}|\left(\frac{1}{2}(1-a)D_n^2-a\log\frac{e}{a}\right)\right),\nonumber\\
\end{IEEEeqnarray}
where $(d)$ follows because 
\begin{IEEEeqnarray}{rCl}
	{|\mathcal{M}| \choose (1-a)|\mathcal{M}|}
	&\leq & \left(\frac{e}{a}\right)^{a|\mathcal{M}|},
	\end{IEEEeqnarray}
where the last inequality follows from \cite{Link2}. Thus, in summary, we get:
\begin{IEEEeqnarray}{rCl}
	\mathbb{P}\left(\mathcal{B}_1\cup\mathcal{B}_2\right) &\leq& 2\exp\left(-|\mathcal{M} \| \mathcal{K}|\left(\frac{1}{2}(1-a)D_n^2-a\log\frac{e}{a}\right)\right)\\
	&\le& \exp\left(-|\mathcal{M} \| \mathcal{K}|\frac{1}{4\log n}\right)\rightarrow 0.
	\label{event0}
\end{IEEEeqnarray}
for sufficiently large $n$.

\underline{\textit{Step 2: Probability of Satisfying the Reliability Condition}}:

 Now, define the following event:
 \begin{IEEEeqnarray}{rCl}
	\mathcal{G} =\left\{\forall s\in\{1,2\},k\in\mathcal{K}\colon \bar{P}_{\mathrm{e}}(s,k)\leq  \mathbb{E}_{\mathcal{C}_s}\left[\bar{P}_{\mathrm{e}}(s,k)\right] +n^{-\frac{1}{6}}\right\}.\label{step201}
	\end{IEEEeqnarray}
Using the McDiarmid's theorem, we shall prove the following lemma, which can be thought as dual to Lemma \ref{le:jingbo},
\begin{lemma}[Concentration of average error probability around its expectation]\label{le:aep-con}
For all $k\in\mathcal{K}$ and any $\Delta>0$, we have
\begin{align}
\mathbb{P}\left(\bar{P}_e(s,k)> \mathbb{E}[\bar{P}_e(s,k)]+\Delta\right)\le \exp_e(-2|\mathcal{M}|\Delta^2).
\end{align}
\end{lemma}

This lemma with $\Delta=n^{-\frac{1}{6}}$ and the union bound imply
\begin{equation}
\mathbb{P}[\mathcal{G}^c]\le 2|\mathcal{K}|\exp(-2|\mathcal{M}|n^{-\frac{1}{3}})\rightarrow 0,\label{step202}
\end{equation}
for sufficiently large $n$, since $\log|\mathcal{K}|=O(\sqrt{n})$.

\begin{IEEEproof}[Proof of Lemma \ref{le:aep-con}]
Note that for a random codebook $\mathcal{C}_s$, the random variable $\bar{P}_{\mathrm{e}}(s,k)$ is a function of $|\mathcal{M}|$ independent random variables  $X_{s,1,k}^n,\cdots,X^n_{s,|\mathcal{M}|,k}$.   In this case, $\bar{P}_{\mathrm{e}}(s,k)$ can be written as follows:
\begin{align}
\bar{P}_{\mathrm{e}}(s,k)&=1-\bar{P}_{\mathrm{c}}(s,k)\\
                                       &=1-\frac{1}{|\mathcal{M}|}\sum_{y^n}\max_mW_s^n\big(y^n\big|X^n_{s,{m},k}\big)
\end{align}
where $\bar{P}_{\mathrm{c}}(s,k)$ is the probability of correct decoding. Define:
\[
g(\bx_{s,1,k},\cdots,\bx_{1,\mathcal{M},k})\triangleq  \frac{1}{|\mathcal{M}|}\sum_{y^n}\max_mW_s^n\left(y^n|x^n_{s,m  ,k}\right).
\]
We next show that $g$ satisfies the bounded difference property in~\eqref{eqn:bnd-diff}, so $\bar{P}_e(s,k)$   satisfies the bounded difference property as well. Let $(x^n_{s,1,k},\cdots,x^n_{s,|\mathcal{M}|,k})$ and $(\bar{x}^n_{s,1,k},\cdots,\bar{x}^n_{s,|\mathcal{M}|,k})$ be two sequences differing only in the $i$-th coordinate, that is


\begin{align}
\bar{x}^n_{s,m,k} = \left\{ \begin{array}{cc}
x^n_{s,m,k} & \mbox{if } m\neq i \\
\mbox{arbitrary}& \mbox{if } m=  i 
\end{array}  \right.
\end{align}
Let the ML solution be defined as
\begin{equation}
\hat{m}_{y^n}\triangleq\argmax_m\; W^n_s(y^n|x^n_{s,m,k}). \label{eqn:ML}
\end{equation}
We should show that 
$g(\bx_{s,1,k},\ldots,\bx_{s,|\mathcal{M}|,k})-g(\bar{x}^n_{s,1,k},\ldots,\bar{x}^n_{s,|\mathcal{M}|,k})$ is bounded. Consider,
\begin{IEEEeqnarray}{rCl}
	&& g(\bx_{s,1,k},\ldots,\bx_{s,|\mathcal{M}|,k})-g(\bar{x}^n_{s,1,k},\ldots,\bar{x}^n_{s,|\mathcal{M}|,k})  \nonumber\\
	&&\hspace{1cm}= \frac{1}{|\mathcal{M}|}\sum_{y^n}\max_{m}W_s^n\left(y^n|x^n_{s,{m},k}\right)- \frac{1}{|\mathcal{M}|}\sum_{y^n}\max_mW_s^n\left(y^n|\bar{x}^n_{s,{m},k}\right)\\
	&&\hspace{1cm}= \frac{1}{|\mathcal{M}|}\sum_{y^n}W_s^n\left(y^n|x^n_{s,\hat{m}_{y^n},k}\right)- \frac{1}{|\mathcal{M}|}\sum_{y^n}\max_mW_s^n\left(y^n|\bar{x}^n_{s,{m},k}\right)\label{eqn:y-0}\\
	&&\hspace{1cm} \le\frac{1}{|\mathcal{M}|}\sum_{y^n}\left(W_s^n\left(y^n|x^n_{s,\hat{m}_{y^n},k}\right)-W_s^n\left(y^n|\bar{x}^n_{s,\hat{m}_{y^n},k}\right)\right)\label{eqn:y-1}\\
	&&\hspace{1cm} \le\frac{1}{|\mathcal{M}|}\sum_{y^n:\hat{m}_{y^n}=i}W_s^n\left(y^n|x^n_{s,\hat{m}_{y^n},k}\right)\label{eqn:y-2}\\
	&&\hspace{1cm}\leq \frac{1}{|\mathcal{M}|},\label{eqn:y-3}
\end{IEEEeqnarray}
where
\begin{itemize}
\item in~\eqref{eqn:y-0}, we used the definition of $\hat{m}_{y^n}$ in \eqref{eqn:ML};
\item in~\eqref{eqn:y-1} follows from the trivial inequality  $\max_m W_s^n (y^n|\bar{x}^n_{s,{m},k})\ge W_s^n (y^n|\bar{x}^n_{s,\hat{m}_{y^n},k})  $;
\item in~\eqref{eqn:y-2} follows because for $\hat{m}_{y^n}\neq i$, $\bx_{s,\hat{m}_{y^n},k}=\bar{x}^n_{s,\hat{m}_{y^n},k}$. 
\end{itemize}

Inequality \eqref{eqn:y-3} implies that $\bar{P}_e(s,k)$ has the bounded difference property with $d_i=\frac{1}{|\mathcal{M}|}$. Finally, the implication of the McDiarmid's theorem completes the proof.
\end{IEEEproof}

\underline{\textit{Step 3: Expurgation}}:
Combining \eqref{event0} and \eqref{step202}, we get:
\begin{IEEEeqnarray}{rCl}
	\mathbb{P}\left[\mathcal{G}^c \cup\mathcal{B}_1\cup\mathcal{B}_2\right] \le \mathbb{P}(\mathcal{G}^c)+\mathbb{P}\left(\mathcal{B}_1\cup\mathcal{B}_2\right) \rightarrow 0.
	\end{IEEEeqnarray}
Thus, there exists a codebook $\mathcal{C}_s^*\in \mathcal{G}\cap \mathcal{B}_1^c\cap  \mathcal{B}_2^c$.

For the codebook $\mathcal{C}_s^*$, for all $s\in\{1,2\}$ and $k\in\mathcal{K}$, we have:
\begin{IEEEeqnarray}{rCl}
	\bar{P}_{\mathrm{e}}(s,k)\leq \mathbb{E}\left[ \bar{P}_{\mathrm{e}}(s,k) \right]+n^{-\frac{1}{6}}=O\left(n^{-\frac{1}{6}}\right),
\end{IEEEeqnarray} 
where we used the bound in Lemma~\ref{PPM-thm}.

 We then remove $a|\mathcal{M}|$ codewords from the codebook $\mathcal{C}_s^*$ with largest $P_{\mathrm{e}}(s,m,k)$ for all $k\in\mathcal{K}$ and $s\in\{1,2\}$ from the subcodebook $\{ X_s^n(m,k)\colon m\in\mathcal{M} \}$ to get  new codebook $\{ X_s^n(m,k)\colon m\in\mathcal{M}_{s,k} \}$ such that $|\mathcal{M}_{s,k}|=(1-a)|\mathcal{M}|$. For these codebooks, the maximum error probability is given by (recall that $a=n^{-\frac{1}{12}}$):
 \begin{IEEEeqnarray}{rCl}
 	\max_m \;P_{\mathrm{e}}(m,s,k)\le \frac{1}{a}O\left(n^{-\frac{1}{6}}\right)=O\left(n^{-\frac{1}{12}}\right)=o(1).
 	\end{IEEEeqnarray}	

Further, \eqref{event0} implies that
\[
d_{\TV}(Q_{\mathcal{M}_s}^n,Q_{s}^{\times n})\le D_n=o(1).
\]
This completes the proof of existence of a codebook with desired properties. 

\section{Conclusion of Proof of Theorem~\ref{key-thm}}\label{key-thm-proof}

Using Lemma \ref{lem4}, if the constraints \eqref{lem-ineq1} and \eqref{lem4-ineq} hold, then there exists a codebook such that the maximum error probability and the following are satisfied:
\begin{align}
d_{\TV}(Q_1^n,Q_2^n)&\leq d_{\TV}(Q_1^n,Q_1^{\times n})+d_{\TV}(Q_1^{\times n},Q_2^{\times n})+d_{\TV}(Q_2^n,Q_2^{\times n})\\
&\stackrel{(a)}{\leq} 2\Phi\left(\frac{1}{2}\sqrt{\Omega(\gamma_1,\gamma_2)}\right)-1+O\left(\frac{1}{\sqrt{n}}\right)+o(1)\\
&\stackrel{(b)}{\leq} \delta +
o(1)
,
\end{align}
where $(a)$ follows from \eqref{lem4-ineq2}  and Lemma~\ref{lem-ach} and $(b)$ follows from \eqref{omega-ineq}. 

Dividing both sides of \eqref{lem-ineq1} and \eqref{lem4-ineq} by $\sqrt{n}$ and letting $n\to \infty$, we get:

\begin{align}
L+L_K \geq (1+\kappa)\cdot \max_{s\in\{1,2\}}\gamma_s
D(W_s \| Q_0|\bar{P}_s),\label{step210}
\end{align}
and 
\begin{IEEEeqnarray}{rCl}
L\leq \min_{s\in\{1,2\}}\gamma_s
D(W_s \| Q_0|\bar{P}_s),\label{step211}
\end{IEEEeqnarray}
where we have used the approximation $I(P_{s,n},W_s)=\frac{\gamma_s}{\sqrt{n}}D(W_s \| Q_0|\bar{P}_s)+o\left(\frac{1}{\sqrt{n}}\right)$. Performing Fourier-Motzkin elimination in \eqref{step210}--\eqref{step211} and letting $\kappa\to 0$, concludes the proof of the theorem.

\section{Solution of the Optimization Problem \eqref{key-ineq1}}\label{cor-proof}
Define:
\begin{IEEEeqnarray}{rCl}
 \bar{\chi}_{2,s}\triangleq  \chi_2(\tilde{Q}_s\|Q_0),\qquad s\in\{1,2\},
\end{IEEEeqnarray}
 $\bar{\rho}\triangleq-2\rho(\tilde{Q}_1,\tilde{Q}_2,Q_0)$ and $\bar{\delta}\triangleq \left(2\Phi^{-1}\left(\frac{1+\delta}{2}\right)\right)^2$. Also, recall the definition of $\bar{\mathbb{D}}_s$ from \eqref{step1000}.
The inequality \eqref{omega-ineq} can be written as follows:
\begin{align}
\bar{\chi}_{2,1}\gamma_1^2+\bar{\chi}_{2,2}\gamma_2^2+\bar{\rho}\gamma_1\gamma_2\leq \bar{\delta}.
\end{align}
The optimization problem in \eqref{key-ineq1} is given by the following:
\begin{align}
&\max_{\gamma_1,\gamma_2}\;\;\min_{s} \gamma_s\bar{\mathbb{D}}_s,\\
&\hspace{0cm}\text{s.t.}:\hspace{0.5cm}\bar{\chi}_{2,1}\gamma_1^2+\bar{\chi}_{2,2}\gamma_2^2+\bar{\rho}\gamma_1\gamma_2\leq \bar{\delta}.
\end{align}
The above optimization problem is equivalently written as:
\begin{IEEEeqnarray}{rCl}
&&	\max_{\gamma_1,\gamma_2}\;\;\min_{0\leq \xi\leq 1}\;\; \xi \gamma_1\bar{\mathbb{D}}_1+(1-\xi)\gamma_2\bar{\mathbb{D}}_2 \\
&&\hspace{0cm}\text{s.t.}:\hspace{0.5cm}\bar{\chi}_{2,1}\gamma_1^2+\bar{\chi}_{2,2}\gamma_2^2+\bar{\rho}\gamma_1\gamma_2\leq \bar{\delta}.\label{cons-min}
	\end{IEEEeqnarray}
Since the above optimization problem is convex, we have:
\begin{IEEEeqnarray}{rCl}
	&&	\min_{0\leq \xi\leq 1}\;\;\max_{\gamma_1,\gamma_2}\;\; \xi \gamma_1\bar{\mathbb{D}}_1+(1-\xi)\gamma_2\bar{\mathbb{D}}_2 \\*
	&&\hspace{0cm}\text{s.t.}:\hspace{0.5cm}\bar{\chi}_{2,1}\gamma_1^2+\bar{\chi}_{2,2}\gamma_2^2+\bar{\rho}\gamma_1\gamma_2\leq \bar{\delta}.
\end{IEEEeqnarray}
Define  
\begin{align}
 G  &\triangleq \xi \gamma_1\bar{\mathbb{D}}_1+(1-\xi)\gamma_2\bar{\mathbb{D}}_2-\lambda\left(\bar{\chi}_{2,1}\gamma_1^2+\bar{\chi}_{2,2}\gamma_2^2+\bar{\rho}\gamma_1\gamma_2- \bar{\delta}\right).
\end{align}
We calculate the derivative of $G$ with respect to $\gamma_1$ and $\gamma_2$ and let it be zero to get the optimal values $\gamma_1^*$ and $\gamma_2^*$. Thus, we obtain 
\begin{align}
2\bar{\chi}_{2,1}\gamma_1^*+\bar{\rho}\gamma_2^*&=\frac{\xi \bar{\mathbb{D}}_1}{\lambda},\\*
2\bar{\chi}_{2,2}\gamma_2^*+\bar{\rho}\gamma_1^*&=\frac{(1-\xi) \bar{\mathbb{D}}_2}{\lambda}.
\end{align}
This yields:
\begin{IEEEeqnarray}{rCl}
	\gamma_1^* &=& \frac{2\bar{\chi}_{2,2}\xi \bar{\mathbb{D}}_1-\bar{\rho}(1-\xi)\bar{\mathbb{D}}_2}{\lambda (4\bar{\chi}_{2,1}\bar{\chi}_{2,2}-\bar{\rho}^2)}\triangleq \frac{\bar{\gamma}_1}{\lambda},\\
	\gamma_2^* &=& \frac{2\bar{\chi}_{2,1}(1-\xi) \bar{\mathbb{D}}_2-\xi \bar{\rho}\bar{\mathbb{D}}_1}{\lambda (4\bar{\chi}_{2,1}\bar{\chi}_{2,2}-\bar{\rho}^2)}\triangleq \frac{\bar{\gamma}_2}{\lambda}.
	\end{IEEEeqnarray}
From \eqref{cons-min}, we know that
\begin{align}
\lambda \geq \sqrt{\frac{\bar{\chi}_{2,1}\bar{\gamma}_1^2+\bar{\chi}_{2,2}\bar{\gamma}_2^2+\bar{\rho}\bar{\gamma}_1\bar{\gamma}_2}{\bar{\delta}}}.
\end{align}
Thus, we have:
\begin{IEEEeqnarray}{rCl}
	\min_{0\leq \xi\leq 1}\;\max_{\gamma_1,\gamma_2}\;\; \xi \gamma_1\bar{\mathbb{D}}_1+(1-\xi)\gamma_2\bar{\mathbb{D}}_2 &&\hspace{0.3cm}=\min_{0\leq \xi\leq 1}\; \sqrt{\frac{\bar{\delta}}{\bar{\chi}_{2,1}\bar{\gamma}_1^2+\bar{\chi}_{2,2}\bar{\gamma}_2^2+\bar{\rho}\bar{\gamma}_1\bar{\gamma}_2}}\left(\xi \bar{\gamma}_1\bar{\mathbb{D}}_1+(1-\xi)\bar{\gamma}_2\bar{\mathbb{D}}_2\right).\nonumber\\
	&& \hspace{0.3cm}=\min_{0\leq \xi\leq 1}\; \sqrt{\frac{4\bar{\delta}\left((1-\xi)^2\bar{\chi}_{2,1}\bar{\mathbb{D}}_2^2-\xi(1-\xi)\bar{\rho} \bar{\mathbb{D}}_1\bar{\mathbb{D}}_2+\xi ^2\bar{\chi}_{2,2} \bar{\mathbb{D}}_1^2 \right)}{\left(4\bar{\chi}_{2,1}\bar{\chi}_{2,2}-\bar{\rho}^2\right)}}\nonumber\\  
	&& \hspace{0.3cm}=\left\{\begin{array}{ll}
	                               \bar{\mathbb{D}}_1\bar{\mathbb{D}}_2\sqrt{\dfrac{\bar{\delta}}{\bar{\chi}_{2,1}\bar{\mathbb{D}}_2^2+\bar{\rho}\bar{\mathbb{D}}_1\bar{\mathbb{D}}_2+\bar{\chi}_{2,2} \bar{\mathbb{D}}_1^2}};&\!\!\!\bar{\rho}>-2\min\left(\frac{\bar{\chi}_{2,1}\bar{\mathbb{D}}_2}{\bar{\mathbb{D}}_1},\frac{\bar{\chi}_{2,2}\bar{\mathbb{D}}_1}{\bar{\mathbb{D}}_2}\right)\\
	                            \sqrt{\dfrac{4\bar{\delta}\min\left(\bar{\chi}_{2,1}\bar{\mathbb{D}}_2^2,\bar{\chi}_{2,2}\bar{\mathbb{D}}_1^2\right)}{4\bar{\chi}_{2,1}\bar{\chi}_{2,2}-\bar{\rho}^2}}; &\!\!\! \bar{\rho}<-{2}\min\left(\frac{\bar{\chi}_{2,1}\bar{\mathbb{D}}_2}{\bar{\mathbb{D}}_1},\frac{\bar{\chi}_{2,2}\bar{\mathbb{D}}_1}{\bar{\mathbb{D}}_2}\right)
	       \end{array} \right.                  
\end{IEEEeqnarray}
This completes the proof.

\section{Proof of \eqref{step3000} for the Example of Fig.~\ref{figure5}}\label{new-ex-proof}
First, notice that for the proposed example, we have
\begin{IEEEeqnarray}{rCl}
	Q_{1,n}(0) = \frac{1-\mu_n}{2},\;\;Q_{1,n}(1)=\frac{1}{2},\;\; Q_{1,n}(2)=\frac{\mu_n}{2},
	\end{IEEEeqnarray}
and 
\begin{IEEEeqnarray}{rCl}
	Q_{2,n}(0) = \frac{1}{2},\;\;Q_{1,n}(1)=\frac{1-\mu_n}{2},\;\; Q_{1,n}(2)=\frac{\mu_n}{2},
\end{IEEEeqnarray}
where $\mu_n=\frac{\gamma}{\sqrt{n}}$ for some positive $\gamma$ such that
\begin{align}
\gamma \leq 2\sqrt{\ln \frac{1}{1-\delta}}.
\end{align}
Now, consider the following upper bound on total variation distance in terms of Bhattacharyya distance \cite[Lemma 3.3.9]{Reis}
\begin{IEEEeqnarray}{rCl}
	d_{\TV}(Q_1^{\times n},Q_2^{\times n})\leq \sqrt{1-F(Q_1^{\times n},Q_2^{\times n})^2}.\label{step2000}
	\end{IEEEeqnarray}
Since $Q_1^{\times n}$ and $Q_{2}^{\times n}$ are product distributions,  one can show that:
\begin{IEEEeqnarray}{rCl}
	F(Q_1^{\times n},Q_2^{\times n}) = F(Q_{1,n},Q_{2,n})^n. \label{step2001}
	\end{IEEEeqnarray}
Calculation of the Bhattacharyya distance for two pmfs $Q_{1,n}$ and $Q_{2,n}$ yields the following:
\begin{IEEEeqnarray}{rCl}
F(Q_{1,n},Q_{2,n}) &=&\frac{1}{2}\left(2\sqrt{1-\mu_n}+\mu_n\right)\\	
&=& 1-\frac{1}{8}\mu_n^2+o(\mu_n^2)\label{step2002}\\
&=& 1-\frac{\gamma^2}{8n}+o\left(\frac{1}{n}\right),
	\end{IEEEeqnarray}
where \eqref{step2002} follows from Taylor series $\sqrt{1-x}= 1-\frac{x}{2}-\frac{x^2}{8}+o(x^2)$ as $x\to 0$.
Combining \eqref{step2000}, \eqref{step2001} and \eqref{step2002}, we have the following upper bound on total variation distance:
\begin{IEEEeqnarray}{rCl}
	d_{\TV}(Q_1^{\times n},Q_2^{\times n})& \leq& \sqrt{1-\left(1-\frac{\gamma^2}{8n}+o\left(\frac{1}{n}\right)\right)^{2n}}\\
	&=&\sqrt{1-\exp\left(-\frac{\gamma^2}{4}\right)}+o(1)\label{step2003}\\
	&\leq &\delta+o(1),
	\end{IEEEeqnarray}
where \eqref{step2003} follows $\left(1-\frac{x}{2n}\right)^{2n}\to \exp(-x)$ as $n\to\infty$. 

To find a lower bound on the optimal throughput, one can repeat the same steps as in Appendices~\ref{lem-key-proof}, \ref{rate-analysis}, \ref{existence2-proof} and get similar results with the following approximations:
\begin{IEEEeqnarray}{rCl}
	I(P_s,W_s) &=& \frac{1}{2}h_{\text{b}}(\mu_n)\\
	&=& \frac{\sqrt{\ln \frac{1}{1-\delta}}}{2}\cdot \frac{\log n}{\sqrt{n}}+o\left(\frac{\log n}{\sqrt{n}}\right),
	\end{IEEEeqnarray}
and 
\begin{IEEEeqnarray}{rCl}
	\sum_{x}\sum_y P_{s,n}(x)W_s(y|x)\left(\log \frac{W_s(y|x)}{Q_{s,n}(y)}\right)^2&=& \frac{1}{2}\left(\mu_n\left(\log \mu_n\right)^2+(1-\mu_n)\left(\log (1-\mu_n)\right)^2\right)\\
	&=& \frac{\sqrt{\ln \frac{1}{1-\delta}}}{4}\cdot \frac{\log^2 n}{\sqrt{n}}+o\left(\frac{\log^2 n}{\sqrt{n}}\right).
	\end{IEEEeqnarray}
Thus, Lemmas~\ref{PPM-thm} and \ref{lem4} state that if we have
\begin{IEEEeqnarray}{rCl}
	\log |\mathcal{M}| \leq \frac{\sqrt{\ln \frac{1}{1-\delta}}}{2}\cdot \sqrt{n}\log n+o\left(\sqrt{n}\log n\right),
	\end{IEEEeqnarray}
then the desired constraints on the maximum error probability and the (total variation) covertness criterion are satisfied.
This completes the proof. 

\section{Proof of  Lemma~\ref{lem2}}\label{lem2-proof}
Let $\mathcal{P}^n$ be the set of all types $\pi_X$ over $\mathcal{X}^n$ and $\mathcal{C}_{s,\pi_X}$ be the sub-codebook  of  $\mathcal{C}_{s}$ consisting of all codewords with the same type $\pi_X$.  Notice that \begin{align}\mathcal{C}_s=\bigcup_{\pi_X\in\mathcal{P}^n}\mathcal{C}_{s,\pi_X}.\end{align} 
 Let $P_{X_s^n}$ be the uniform distribution over the codebook $\mathcal{C}_s= \{x_s^n(m,k):m\in\mathcal{M},k\in\mathcal{K}\}$, i.e., for every $x^n\in\mathcal{X}^n$:
\begin{align}
P_{X_s^n}(x^n)= \left\{ \begin{array}{cc}
\frac{1}{|\mathcal{M} |\cdot |\mathcal{K} |} & x^n \in\mathcal{C}_s \\
0 & \mbox{ else }
\end{array} \right.
\end{align}	
In the following, we first consider false alarm probability and next the missed detection probability. Consider the false alarm probability as follows:
\begin{align}
\alpha_n &=  \mathbb{P}_{Q_1^n}\left(\sum_{i=1}^n\mathcal{T}_{\text{\test}}(Y_i)\leq  \tau\right)\\
&= \sum_{y^n} Q_1^n(y^n) \mathbbm{1} \left\{\sum_{i=1}^n\mathcal{T}_{\text{\test}}(y_i)\leq  \tau\right\}\\
&= \sum_{y^n}\bigg( \sum_{x^n  \in \mathcal{C}_{1}} P_{X_1^n}(x^n)W_1^n(y^n|x^n) \bigg) \mathbbm{1} \left\{\sum_{i=1}^n\mathcal{T}_{\text{\test}}(y_i)\leq \tau\right\}\\
&= \sum_{x^n\in\mathcal{C}_{1}}P_{X_1^n}(x^n) \sum_{y^n } W_1^n(y^n|x^n)  \mathbbm{1} \left\{\sum_{i=1}^n\mathcal{T}_{\text{\test}}(y_i)\leq \tau\right\}\\
&= \sum_{\pi_X}\sum_{x^n\in\mathcal{C}_{1,\pi_X}}P_{{X_1^n}}(x^n) \mathbb{P}_{W_1^n(\cdot|x^n)}\left( \sum_{i=1}^n\mathcal{T}_{\text{\test}}(Y_i)\leq \tau \right)
\end{align}
Since  $ \mathbb{P}_{W_1^n(\cdot|x^n)}\left(\sum_{i=1}^n\mathcal{T}_{\text{\test}}(Y_i)\leq \tau \right) $ remains the same for each $x^n$ with the same type, and $\{P_{{X_1^n}} \{  \mathcal{C}_{1,\pi_X} \}\}_{\pi_X\in\mathcal{P}^n}$ is a probability distribution,  we have 
\begin{align}
\alpha_n \le \max_{\pi_X}\;  \mathbb{P}_{W_1^n(\cdot|x_1^n)}\left( \sum_{i=1}^n\mathcal{T}_{\text{\test}}(Y_i)\leq \tau \right)\label{pr1}, 
\end{align}
where $x_1^n$ above refers to any vector with type $\pi_X$.  
Denoting the maximizing type in \eqref{pr1} by $\pi^*$ and letting $x^{*n}$ be any vector with type $\pi^*$, \eqref{pr1} can be written as follows:
\begin{IEEEeqnarray}{rCl}
	\alpha_n 
	&\le&\mathbb{P}_{W_1^n(\cdot |x^{*n})}\left(\sum_{i=1}^n\mathcal{T}_{\text{\test}}(Y_i)\leq \tau \right).\label{alpha-upp}
	\end{IEEEeqnarray}
  Notice that  $\pi^*$ relates to the Hamming weight of the codeword $x^{*n}$ as follows:
  \begin{IEEEeqnarray}{rCl}
  	\pi^*(x)=\left\{\begin{array}{ll} \frac{\w(x^{*n})}{n} & x=1\\ 1-\frac{\w(x^{*n})}{n} & x=0\end{array} \right.
  	\end{IEEEeqnarray}
 We also define
\begin{IEEEeqnarray}{rCl}
	\mu_{1,n}\triangleq \pi^*(1)= \frac{\w(x^{*n})}{n}.
\end{IEEEeqnarray}
Since the channel is memoryless, $\left\{\mathcal{T}_{\text{test}}(Y_i)\right\}_{i=1}^n$ are mutually independent so Theorem~\ref{BE-thm} can be applied to upper bound the probability in \eqref{alpha-upp}. For every $x^{*n}$ with type $\pi^*$, we calculate $\sum_{i=1}^n\mathbb{E}\left[\mathcal{T}_{\text{test}}(Y_i) \right]$ and $\sum_{i=1}^n\mathbb{V}\left[\mathcal{T}_{\text{test}}(Y_i) \right]$ in the following. First, we calculate the expectation as follows,
\begin{IEEEeqnarray}{rCl}
	\sum_{i=1}^n\mathbb{E}_{W_1(\cdot |x^*_{i})}\left[\mathcal{T}_{\test}(Y_i)\right]
	&=&\sum_{i=1}^n\mathbb{E}_{W_1(\cdot |x_{i}^*)}\left[\frac{\tilde{Q}_{1}(Y_i)-\tilde{Q}_{2}(Y_i)}{Q_{0}(Y_i)}\right]\\
	&=&\sum_{\substack{i:x_{i}^*=1}}\mathbb{E}_{\tilde{Q}_{1}}\left[\frac{\tilde{Q}_{1}(Y_i)-\tilde{Q}_{2}(Y_i)}{Q_{0}(Y_i)}\right]\label{step31}\\
	&=&n\mu_{1,n}\sum_y \frac{\tilde{Q}_{1}(y)(\tilde{Q}_{1}(y)-\tilde{Q}_{2}(y))}{Q_{0}(y)}\\
	&=&n\mu_{1,n}\sum_y \frac{(\tilde{Q}_{1}(y)-Q_0(y))(\tilde{Q}_{1}(y)-\tilde{Q}_{2}(y))}{Q_{0}(y)}\\
	&=& n\mu_{1,n} \D_{1},\label{def1}
\end{IEEEeqnarray}
where \eqref{step31} follows because when $x_{i}^*=0$, we have $W_1(Y_i|x_{i}^*)=Q_0(Y_i)$ and the expectation in the summand is equal to zero. Next, we calculate the variance as follows
\begin{IEEEeqnarray}{rCl}
 \sum_{i=1}^n\mathbb{V}\left[\mathcal{T}_{\text{test}}(Y_i)\right]
	&=& \sum_{i=1}^n  \mathbb{E}\left[\left(\mathcal{T}_{\text{test}}(Y_i)\right)^2\right]-\sum_{i=1}^n\left(\mathbb{E}\left[\mathcal{T}_{\text{test}}(Y_i)\right]\right)^2 ,\label{step32}
\end{IEEEeqnarray}
Now, we calculate each expectation term in \eqref{step32} as follows,
 \begin{IEEEeqnarray}{rCl}
\sum_{i=1}^n \mathbb{E}\left[\left(\mathcal{T}_{\text{test}}(Y_i)\right)^2\right] &=&  \sum_{i=1}^n  \mathbb{E}_{W_1(\cdot |x_{i}^*)}\left[\left(\mathcal{T}_{\text{test}}(Y_i)\right)^2\right]	\\
&=& \sum_{i:x_{i}^*=1} \mathbb{E}_{\tilde{Q}_1}\left[\left(\frac{\tilde{Q}_{1}(Y_i)-\tilde{Q}_{2}(Y_i)}{Q_{0}(Y_i)}\right)^2\right]+\sum_{i:x_{i}^*=0} \mathbb{E}_{Q_0}\left[\left(\frac{\tilde{Q}_{1}(Y_i)-\tilde{Q}_{2}(Y_i)}{Q_{0}(Y_i)}\right)^2\right]\\
&=& n\mu_{1,n}\sum_y\frac{\tilde{Q}_1(y)\big(\tilde{Q}_{1}(y)-\tilde{Q}_{2}(y)\big)^2}{Q_{0}^2(y)}+n(1-\mu_{1,n})\Delta,\nonumber\\\label{step102}
	\end{IEEEeqnarray}
and 
\begin{IEEEeqnarray}{rCl}
\sum_{i=1}^n\left(\mathbb{E}\left[\mathcal{T}_{\text{test}}(Y_i)\right]\right)^2&=& \sum_{i=1}^n\left(\mathbb{E}_{W_1(\cdot |x_{i}^*)}\left[\mathcal{T}_{\text{test}}(Y_i)\right]\right)^2\\
&=& \sum_{i=1}^n\left(\mathbb{E}_{W_1(\cdot|x_{i}^*)}\left[\frac{\tilde{Q}_{1}(Y_i)-\tilde{Q}_{2}(Y_i)}{Q_{0}(Y_i)}\right]\right)^2\\
&=&\sum_{i:x_{i}^*=1}\left(\mathbb{E}_{\tilde{Q}_1}\left[\frac{\tilde{Q}_{1}(Y_i)-\tilde{Q}_{2}(Y_i)}{Q_{0}(Y_i)}\right]\right)^2+ \sum_{i:x_{i}^*=0}\left(\mathbb{E}_{Q_0}\left[\frac{\tilde{Q}_{1}(Y_i)-\tilde{Q}_{2}(Y_i)}{Q_{0}(Y_i)}\right]\right)^2 \\[1ex]
&=&n\mu_{1,n}\left(\sum_y\frac{\tilde{Q}_1(y)\big(\tilde{Q}_1(y)-\tilde{Q}_2(y)\big)}{Q_0(y)}\right)^2+n(1-\mu_{1,n})\left(\sum_y\big(\tilde{Q}_1(y)-\tilde{Q}_2(y)\big)\right)^2\\
&=&n\mu_{1,n}\left(\sum_y\frac{\tilde{Q}_1(y)\big(\tilde{Q}_1(y)-\tilde{Q}_2(y)\big)}{Q_0(y)}\right)^2\\
&=&n\mu_{1,n}\left(\sum_y\frac{(\tilde{Q}_1(y)-Q_0(y))(\tilde{Q}_1(y)-\tilde{Q}_2(y))}{Q_0(y)}\right)^2\\
&=&n\mu_{1,n}\D_1^2.\label{step103}
	\end{IEEEeqnarray}

Combining \eqref{step32}, \eqref{step102} and \eqref{step103}, we get 
\begin{IEEEeqnarray}{rCl}
	\mathbb{V}\left[Z_n(x^{*n})\right]
	&=&n\Bigg(\mu_{1,n} \sum_y \frac{\tilde{Q}_{1}(y)\big(\tilde{Q}_{1}(y)-\tilde{Q}_{2}(y)\big)^2}{Q_{0}^2(y)}+(1-\mu_{1,n})\Delta-\mu_{1,n}\D_{1}^2\Bigg)\\
	&\triangleq &n\V_{1,n}.\label{step41}
\end{IEEEeqnarray}
We now bound the sum of the  third absolute moments 
\begin{IEEEeqnarray}{rCl}
	\sum_{i=1}^n \mathbb{E}_{W_1(\cdot|x_i^{*})}\left[ \big|\mathcal{T}_{\test}(Y_i) -\mathbb{E}_{W_1(\cdot|x_i^{*})} [ \mathcal{T}_{\test}(Y_i) ] \big|^3 \right]\label{third1}.
\end{IEEEeqnarray}
We know that $\mathcal{T}_{\test}(y_i)= \frac{\tilde{Q}_1(y_i)-\tilde{Q_2}(y_i)}{Q_0(y_i)}$. Hence,
\begin{align}|\mathcal{T}_{\test}(y_i)|  \le \frac{2}{\min_{y_i} Q_0(y_i)} \triangleq 2/\eta <\infty.\label{t-bound}\end{align}	
By the triangle inequality,
\begin{align}
\big|\mathcal{T}_{\test}(Y_i) -\mathbb{E}  [ \mathcal{T}_{\test}(Y_i) ] \big|\le 4/\eta,\quad\mbox{a.s.}\label{third2}
\end{align}
Combining \eqref{third1} and \eqref{third2}, we obtain
\begin{align}
&\sum_{i=1}^n \mathbb{E}_{W_1(\cdot|x_i^{*})}\left[ \big|\mathcal{T}_{\test}(Y_i) -\mathbb{E}_{W_1(\cdot|x_i^{*})} [ \mathcal{T}_{\test}(Y_i) ] \big|^3\right]\leq \frac{64n}{\eta^3}\triangleq n\T.\label{third-abs5}
\end{align}

Notice that $\V_{1,n}$ in \eqref{step41} is further upper bounded as follows:
\begin{IEEEeqnarray}{rCl}
	\V_{1,n}&\leq& \Bigg(\mu_{1,n} \sum_y \frac{\tilde{Q}_{1}(y)\big(\tilde{Q}_{1}(y)-\tilde{Q}_{2}(y)\big)^2}{Q_{0}^2(y)}+(1-\mu_{1,n})\Delta\Bigg)\\
	&=& \Delta+\mu_{1,n}\Gamma_1\\
	&\leq& \Delta+\mu_{\Hi,n}|\Gamma_1|\\
	&\triangleq &\V_{1,n}^{*}.
\end{IEEEeqnarray}
Thus, from Theorem~\ref{BE-thm}, we get the following:
\begin{IEEEeqnarray}{rCl}
	\hspace{0cm}\mathbb{P}\left(\sum_{i=1}^n\mathcal{T}_{\text{\test}}(Y_i)\leq  \tau\right)&\leq& 	1-\Phi\left(\frac{-\tau+n \mu_{1,n}\D_{1}}{\sqrt{n \V_{1,n}}}\right)+\frac{6\T}{\V_{1,n}^{\frac{3}{2}}\sqrt{n}}.\label{step4}
\end{IEEEeqnarray} 
Combining \eqref{pr1} and \eqref{step4}, we get the following:
\begin{IEEEeqnarray}{rCl}
	\alpha_n \leq \max_{\mu_{1,n}} \;1-\Phi\left(\frac{-\tau+n \mu_{1,n}\D_{1}}{\sqrt{n \V_{1,n}}}\right)+O\left(\frac{1}{\sqrt{n}}\right),\;\;\;\;\;\;\label{step5}
\end{IEEEeqnarray}
where the maximization is over all $\mu_{1,n}$ such that $\mu_{\Lo,n}\leq \mu_{1,n}\leq \mu_{\Hi,n}$.
Now, consider the missed detection probability as follows:
\begin{IEEEeqnarray}{rCl}
	\beta_n &=& \mathbb{P}_{Q_2^n}\left( \sum_{i=1}^n\mathcal{T}_{\test}(Y_i)> \tau \right).
\end{IEEEeqnarray}
Following similar steps leading to \eqref{pr1}, we can write the missed detection probability as:
\begin{IEEEeqnarray}{rCl}
	\beta_n 
	&\leq & \max_{\mu_{2,n}}\;1-\Phi\left(\frac{\tau-n\mu_{2,n}\D_{2}}{\sqrt{n \V_{2,n}}}\right)+\frac{6\T}{\V_{2,n}^{\frac{3}{2}}\sqrt{n}},
	\label{alpha-final}
\end{IEEEeqnarray}
where the maximization is over all $\mu_{2,n}$ such that $\mu_{\Lo,n}\leq \mu_{2,n}\leq \mu_{\Hi,n}$ and we define:
\begin{IEEEeqnarray}{rCl}
	\V_{2,n} &\triangleq& \Bigg(\mu_{2,n} \sum_y \frac{\tilde{Q}_{2}(y)\big(\tilde{Q}_{1}(y)-\tilde{Q}_{2}(y)\big)^2}{Q_{0}^2(y)}+(1-\mu_{2,n})\Delta-\mu_{2,n}\bigg(\sum_y \frac{\tilde{Q}_{2}(y)\big(\tilde{Q}_{1}(y)-\tilde{Q}_{2}(y)\big)}{Q_{0}(y)}\bigg)^2\Bigg).
\end{IEEEeqnarray}
We can further upper bound $\V_{2,n}$ as follows:
\begin{align}
	\V_{2,n}&\leq  \mu_{2,n} \sum_y \frac{\tilde{Q}_{2}(y)\big(\tilde{Q}_{1}(y)-\tilde{Q}_{2}(y)\big)^2}{Q_{0}^2(y)}+(1-\mu_{2,n})\Delta \\
	&= \Delta+ \mu_{2,n}  \Gamma_2\\
	&\leq   \Delta+ \mu_{\Hi,n} |\Gamma_2| \\
	&\triangleq  \V^{*}_{2,n},
\end{align}
 The proof is followed by upper bounding the false alarm and missed detection probabilities using the choice of $\tau$ in \eqref{step15}:
\begin{IEEEeqnarray}{rCl}
	\alpha_n&\leq&\max_{\mu_{1,n}}\;1-\Phi\left(\frac{-\tau+n\mu_{1,n}\D_{1}}{\sqrt{n \V_{1,n}}}\right)+O\left(\frac{1}{\sqrt{n}}\right)\\
	&=& \max_{\mu_{1,n}}\;1-\Phi\left(\frac{-\frac{n\mu_{\Lo,n}}{2}\left(\D_{2}+\D_{1}\right)+n\mu_{1,n}\D_{1}}{\sqrt{n \V_{1,n}}}\right)+O\left(\frac{1}{\sqrt{n}}\right)\\
	&=& 1-\Phi\left(\frac{-\frac{n\mu_{\Lo,n}}{2}\left(\D_{2}+\D_{1}\right)+n\mu_{\Lo,n}\D_{1}}{\sqrt{n \V_{1,n}}}\right)+O\left(\frac{1}{\sqrt{n}}\right)\nonumber\\\\
	&\leq&	1-\Phi\left(\frac{\sqrt{n}\mu_{\Lo,n}\left(\D_{1}-\D_{2}\right)}{2\sqrt{ \V_{1,n}}}\right)+O\left(\frac{1}{\sqrt{n}}\right)\label{step301}\\[1ex]
	&=&1-\Phi\left(\frac{\sqrt{n}\mu_{\Lo,n} \Delta}{2\sqrt{ \V_{1,n}}}\right)+O\left(\frac{1}{\sqrt{n}}\right)\label{step100}\\[1ex]
	&\leq &1-\Phi\left(\frac{\sqrt{n}\mu_{\Lo,n} \Delta}{2\sqrt{ \V^*_{1,n}}}\right)+O\left(\frac{1}{\sqrt{n}}\right)\label{step305}\\[1ex]
	&=& 1-\Phi\left(\frac{\sqrt{n} \mu_{\Lo,n} \Delta}{2\sqrt{ \left(\Delta+\mu_{\Hi,n} |\Gamma_1|\right)}}\right)+O\left(\frac{1}{\sqrt{n}}\right)\\[1.5ex]
	&\leq & 1-\Phi\left(\frac{1}{2}\mu_{\Lo,n}\sqrt{n\Delta} \cdot \left(1-\frac{\mu_{\Hi,n} |\Gamma_1|}{2\Delta}\right)\right)+O\left(\frac{1}{\sqrt{n}}\right)\nonumber\\\label{step6}\\[1ex]
	&\leq & 1-\Phi\left(\frac{1}{2}\mu_{\Lo,n}\sqrt{n\Delta}  \right)+\frac{\sqrt{n}\mu_{\Lo,n}\mu_{\Hi,n}|\Gamma_1|}{4\sqrt{2\pi\Delta}}+O\left(\frac{1}{\sqrt{n}}\right),\label{step16}
\end{IEEEeqnarray}
where
\begin{itemize}
	\item \eqref{step301} follows because assumption \eqref{conv-cons1} implies that $\D_1\geq 0$;
 	\item
	\eqref{step100} follows because $\D_1-\D_2=\Delta$;
	\item \eqref{step305} follows because $\V_{1,n}\leq \V^*_{1,n}$;
	\item \eqref{step6} follows from $\frac{1}{\sqrt{1+x}}\geq 1-\frac{x}{2}$; 
	\item \eqref{step16} follows because $1-\Phi(x-y)\leq 1-\Phi(x)+\frac{y}{\sqrt{2\pi}}$ for all $0<y<x$. 
\end{itemize}
With the choice of $\tau$ in \eqref{step15}, $\beta_n$ in \eqref{step5} can be upper bounded as follows:
\begin{IEEEeqnarray}{rCl}
	\beta_n&\leq & \max_{\mu_{2,n}}\; 1-\Phi\left(\frac{\tau-n \mu_{2,n}\D_{2}}{\sqrt{n \V_{2,n}}}\right)+O\left(\frac{1}{\sqrt{n}}\right)\label{step45}\\
	&=& \max_{\mu_{2,n}}\;1-\Phi\left(\frac{\frac{n\mu_{\Lo,n}}{2}\left(-\D_{2}+\D_{1}\right)+n( \mu_{\Lo,n}-\mu_{2,n})\D_{2}}{\sqrt{n \V_{2,n}}}\right)+O\left(\frac{1}{\sqrt{n}}\right)\\[1.5ex]
	&\leq &   1-\Phi\left(\frac{\frac{n\mu_{\Lo,n}}{2}\left(-\D_{2}+\D_{1}\right)}{\sqrt{n \V_{2,n}}}\right)+O\left(\frac{1}{\sqrt{n}}\right)\label{step300}\\[1.5ex]
	&=&1-\Phi\left(\frac{n\mu_{\Lo,n}\Delta}{2\sqrt{n \V_{2,n}}}\right)+O\left(\frac{1}{\sqrt{n}}\right)\label{step204}\\[1.5ex]
	&\leq &1-\Phi\left(\frac{n\mu_{\Lo,n}\Delta}{2\sqrt{n \V_{2,n}^*}}\right)+O\left(\frac{1}{\sqrt{n}}\right)\label{step302}\\[1.5ex]
	&\leq & 1-\Phi\left(\frac{1}{2}\mu_{\Lo,n}\sqrt{n\Delta}\right)+\frac{\sqrt{n}\mu_{\Lo,n}\mu_{\Hi,n}|\Gamma_2|}{4\sqrt{2\pi\Delta}}+O\left(\frac{1}{\sqrt{n}}\right),\label{step303}
\end{IEEEeqnarray} 
where 
\begin{itemize}
\item \eqref{step300} follows because assumption \eqref{conv-cons1} implies that $\D_2\leq 0$ and hence, $(\mu_{\Lo,n}-\mu_{2,n})\D_2\geq 0$;
\item \eqref{step204} follows because $\D_1-\D_2=\Delta$;
\item \eqref{step302} follows because $\V_{2,n}\leq \V_{2,n}^*$;
\item \eqref{step303} follows because $\V_{2,n}^*=\Delta+\mu_{\Hi,n}|\Gamma_2|$ and also from the fact that $1-\Phi(x-y)\leq 1-\Phi(x)+\frac{y}{\sqrt{2\pi}}$ for all $0<y<x$.
\end{itemize}

This completes the proof of lemma.

\section{Conclusion of Proof of Theorem~\ref{thm-upper}}\label{existence-proof}

 In the following, we first show that there exists a sub-codebook which satisfies \eqref{step42} and its size is at least $\frac{|\mathcal{M}|\cdot |\mathcal{K}|}{\sqrt{n}}$. Define the following set of codewords for some $\gamma>0$:
 \begin{IEEEeqnarray}{rCl}
	\mathcal{D}_s\triangleq\left\{ x_s^n\in\mathcal{C}_s\colon \w(x_s^n)\leq 2\sqrt{\frac{n}{\Delta}}\Phi^{-1}\left(\frac{1+\delta}{2}+\gamma+\frac{E}{\sqrt{n}}\right) \right\},\label{Ds-def}
 \end{IEEEeqnarray}
 where $E$ is chosen such that
 \begin{IEEEeqnarray}{rCl}
 	E> \frac{|\Gamma_1|+|\Gamma_2|}{8\sqrt{2\pi\Delta}}\cdot \left(\Phi^{-1}\left(\frac{1+\delta}{2}+\gamma\right)\right)^2.\label{step13}
 \end{IEEEeqnarray}

 Let $\widehat{Q}_{s}^n$ and $\overline{Q}_{s}^n$ be the induced output distributions for codes $\mathcal{D}_s$ and $\mathcal{C}_s \backslash \mathcal{D}_s$, respectively. The distribution $Q_s^n$ can be written as follows:
 \begin{IEEEeqnarray}{rCl}
 	Q_{s}^n= \xi_s\widehat{Q}_{s}^n+(1-\xi_s)\overline{Q}_{s}^n,
 \end{IEEEeqnarray}
 where $\xi_s\triangleq \frac{|\mathcal{D}_s|}{|\mathcal{M}|\cdot |\mathcal{K}|}$. Without loss of generality, assume that $\xi_1\geq \xi_2$.
 Then, we get the following:
 \begin{IEEEeqnarray}{rCl}	
 	\delta &\geq& d_{\TV}\left(Q_1^n,Q_2^n\right)\\
 	&=&\frac{1}{2} \sum_{y^n}\left|Q_1^n-Q_2^n\right|\\
 	&=&\frac{1}{2}\sum_{y^n}\left|\xi_1\widehat{Q}_1^n+(1-\xi_1)\overline{Q}_1^n-\xi_2\widehat{Q}_2^n-(1-\xi_2)\overline{Q}_2^n\right|\nonumber\\\\
 	&\geq &d_{\TV}\left(\overline{Q}_{1}^n,\overline{Q}_{2}^n\right) -\xi_1  d_{\TV}\left(\widehat{Q}_{1}^n,\overline{Q}_{1}^n\right)-\xi_2 d_{\TV}\left(\widehat{Q}_{2}^n,\overline{Q}_2^n\right),\nonumber\\\label{step11}
 \end{IEEEeqnarray}
 where the last inequality follows from the triangle inequality. 
 We know that for any $x_s^n\in \mathcal{C}_s\backslash \mathcal{D}_s$, 
 \begin{IEEEeqnarray}{rCl}
 	\w(x_s^n)\geq 2\sqrt{\frac{n}{\Delta}}\Phi^{-1}\left(\frac{1+\delta}{2}+ \gamma+\frac{E}{\sqrt{n}}\right),\label{step10}
 \end{IEEEeqnarray}
 Combining \eqref{step42} with \eqref{step10} and considering \eqref{step13}, we can further lower bound the total variation distance as follows:
 \begin{IEEEeqnarray}{rCl}
 	d_{\TV}\left(\overline{Q}_{1}^n,\overline{Q}_{2}^n\right)&\geq& \delta+\frac{2E}{\sqrt{n}}+2\gamma-\frac{\omega_{\Lo,n}\omega_{\Hi,n}(|\Gamma_1|+|\Gamma_2|)}{4n\sqrt{2n\pi\Delta}} +O\left(\frac{1}{\sqrt{n}}\right)\\[1.5ex]
 	&\geq& \delta+2\gamma,
 	\label{step12}
 \end{IEEEeqnarray}
 where the last inequality follows from \eqref{step13}.
Uniting \eqref{step11} and \eqref{step12}   yields 
 \begin{IEEEeqnarray}{rCl}
 	\delta &\geq& \delta+2\gamma-\xi_1-\xi_2.
 \end{IEEEeqnarray}
 If we choose $\gamma=\frac{1}{\sqrt{n}}$, we obtain
 \begin{align}
 \xi_1 +\xi_2\geq \frac{2}{\sqrt{n}}.\label{step400}
 \end{align}
 In summary, from the assumption $\xi_1\geq \xi_2$, we obtain a set of codewords with size \begin{align}
 |\mathcal{D}_1| \geq \frac{|\mathcal{M}|\cdot |\mathcal{K}|}{\sqrt{n}},\label{ineq100}
 \end{align} and the Hamming weight of these codewords is given by 
 \begin{align}
 \w(x_1^n)&\leq 2\sqrt{\frac{n}{\Delta}}\Phi^{-1}\left(\frac{1+\delta}{2}+\frac{E}{\sqrt{n}}+\gamma\right)\\
 &=2\sqrt{\frac{n}{\Delta}}\Phi^{-1}\left(\frac{1+\delta}{2}\right)+O\left(\frac{1}{\sqrt{n}}\right)\\
 &\triangleq \psi(n,\delta)+O\left(\frac{1}{\sqrt{n}}\right).
 \end{align}

Since we have assumed $\xi_1\geq \xi_2$, inequality \eqref{ineq100} can be equivalently written as:
\begin{IEEEeqnarray}{rCl}
	\max_{s}	|\mathcal{D}_s| \geq \frac{|\mathcal{M}|\cdot |\mathcal{K}|}{\sqrt{n}}\label{ineq101}
\end{IEEEeqnarray}

For each $k\in\mathcal{K}$, we denote the sub-codebook of all codewords with the same key $k$ by $\mathcal{C}_s^k$. From the pigeonhole principle, there exists a sub-codebook $\mathcal{C}_s^k$ such that $\max_s |\mathcal{D}_s\cap \mathcal{C}_s^k|\geq \frac{|\mathcal{M}|}{\sqrt{n}}$. Define:
\begin{IEEEeqnarray}{rCl}
	\mathcal{D}_{s,i}^k\triangleq\left\{ x_s^n\in\mathcal{D}_s\cap \mathcal{C}_s^k\colon \w(x_s^n)=i \right\}.\label{def500}
\end{IEEEeqnarray}
The codewords in the sub-codebook $|\mathcal{D}_{s,i}^k|$ have the same type which we denote by $\pi_s^i$. This sub-codebook is  with maximum probability of error not larger than $\epsilon$. Using Fano's inequality, we can write the following set of inequalities for $s\in\{1,2\}$:
\begin{IEEEeqnarray}{rCl}
	(1-\epsilon)\log |\mathcal{D}_{s,i}^k\cap \mathcal{C}_s^k|-1&\leq& I(X_s^n;Y^n)\\
	&\leq & \sum_{t=1}^nI(X_{s,t};Y_t)\\
	&\leq &nI(\pi_s^i,W_s)\\
	&=& iD(\tilde{Q}_{s}\|Q_0).\label{ineq401}
\end{IEEEeqnarray}
Next, we continue to upper bound the size of the message set as follows:
\begin{IEEEeqnarray}{rCl}
	\log \frac{|\mathcal{M}|}{\sqrt{n}}&\stackrel{(a)}{\leq}& \max_s\;\log |\mathcal{D}_s\cap \mathcal{C}_s^k|\\
	&\stackrel{(b)}{=}& \max_s\;\log\left(\sum_{i=0}^{\psi(n,\delta)}|\mathcal{D}_{s,i}^k\cap \mathcal{C}_s^k|\right)\\
	&\stackrel{(c)}{\leq} & \max_s\;\log\left( \sum_{i=0}^{\psi(n,\delta)} 2^{iD(\tilde{Q}_{s}\|Q_0)}\right)\\
	&\leq & \max_s\;\log\left( \psi(n,\delta) 2^{\psi(n,\delta)\cdot D(\tilde{Q}_{s}\|Q_0)}\right)\\
	&\leq & \psi(n,\delta)\cdot  \max_s\; D(\tilde{Q}_{s}\|Q_0)+O(1)\\
	&\leq & 2\sqrt{\frac{n}{\Delta}}\Phi^{-1}\left(\frac{1+\delta}{2}\right)\cdot  \max_s\, D(\tilde{Q}_{s}\|Q_0)+O(1),\label{ineq700}
\end{IEEEeqnarray}
where $(a)$ follows from \eqref{ineq101}, $(b)$ follows from the definition in \eqref{def500}, $(c)$ follows from inequality \eqref{ineq401}.
Dividing both sides of \eqref{ineq700} by $\sqrt{n}$ and taking limits in $n$ completes the proof.

\section{Proof of Corollary~\ref{Gaus-cor}}\label{Gaus-proof}
For the proposed Gaussian setup, the lower bound to $L_{\TV}^*(\epsilon,\delta)$ in Theorem~\ref{PPM-thm} can be generalized for a continuous alphabet. Several steps in the proof of upper bound to $L_{\TV}^*(0,\delta)$ in Appendix~\ref{lem2-proof} remain valid for the Gaussian distribution. The only step that should be refined is bounding the sum of third absolute moments in \eqref{t-bound}. The reason is as follows. In this step of the proof for the discrete memoryless channel, we know that $\eta=\min_y Q_0(y)$ is positive. However, for the Gaussian case, this statement does not hold and $\eta$ is zero. So, in the following, we first bound the sum of third absolute moments for the Gaussian setup. The evaluation of the involved information quantities with a Gaussian distribution will be presented later.

\underline{\textit{Bounding the sum of third absolute moments}}:

Consider the following sum of third absolute moments:

\begin{IEEEeqnarray}{rCl}
	\sum_{i=1}^n \mathbb{E}_{W_1(\cdot|x_i^{*})}\left[ \big|\mathcal{T}_{\test}(Y_i) -\mathbb{E}_{W_1(\cdot|x_i^{*})} [ \mathcal{T}_{\test}(Y_i) ] \big|^3 \right] &=& \sum_{i:x_i^*=0} \mathbb{E}_{W_1(\cdot|x_i^{*})}\left[ \big|\mathcal{T}_{\test}(Y_i) -\mathbb{E}_{W_1(\cdot|x_i^{*})} [ \mathcal{T}_{\test}(Y_i) ] \big|^3 \right] \nonumber\\&&\hspace{1cm}+\sum_{i:x_i^*=1} \mathbb{E}_{W_1(\cdot|x_i^{*})}\left[ \big|\mathcal{T}_{\test}(Y_i) -\mathbb{E}_{W_1(\cdot|x_i^{*})} [ \mathcal{T}_{\test}(Y_i) ] \big|^3 \right] \\
	&=&n(1-\mu_{1,n}) \mathbb{E}_{Q_0}\left[ \big|\mathcal{T}_{\test}(Y) -\mathbb{E}_{Q_0} [ \mathcal{T}_{\test}(Y) ] \big|^3 \right] \nonumber\\&&\hspace{1cm}+n\mu_{1,n} \mathbb{E}_{\tilde{Q}_1}\left[ \big|\mathcal{T}_{\test}(Y) -\mathbb{E}_{\tilde{Q}_1} [ \mathcal{T}_{\test}(Y) ] \big|^3 \right]. \label{third-abs1}
\end{IEEEeqnarray}
We study each expectation term of \eqref{third-abs1}, separately, and show that each expectation term is a constant and does not depend on $n$. First, consider the following term:
\begin{IEEEeqnarray}{rCl}
	\mathbb{E}_{Q_0}\left[ \big|\mathcal{T}_{\test}(Y) -\mathbb{E}_{Q_0} [ \mathcal{T}_{\test}(Y) ] \big|^3 \right] &\stackrel{(a)}{=}& \mathbb{E}_{Q_0}\left[ \big|\mathcal{T}_{\test}(Y)  \big|^3 \right] \label{step700}\\
	&=& \mathbb{E}_{Q_0}\left[ \Bigg| \frac{\tilde{Q}_1(Y)-\tilde{Q}_2(Y)}{Q_0(Y)}  \Bigg|^3 \right]\\
	&=& \frac{1}{\sqrt{2\pi}}\int_{-\infty}^{\infty} \Bigg|\frac{1}{\sigma_1}\exp\left(-\frac{(y-1)^2}{2\sigma_1^2}\right)-\frac{1}{\sigma_2}\exp\left(-\frac{(y-1)^2}{2\sigma_2^2}\right)\Bigg|^3\cdot \sigma_0^2\exp\left(\frac{y^2}{\sigma_0^2}\right)dy\nonumber\\\\
	&\stackrel{(b)}{\leq} & \frac{1}{\sqrt{2\pi}}\int_{-\infty}^{\infty} \frac{1}{\sigma_1^3}\exp\left(-\frac{3(y-1)^2}{2\sigma_1^2}\right)\cdot \sigma_0^2\exp\left(\frac{y^2}{\sigma_0^2}\right)dy+\kappa_1\label{step701}\\
	&=&  \frac{\sigma_0^2}{\sqrt{2\pi}\sigma_1^3}\int_{-\infty}^{\infty} \exp \left(-\frac{3\sigma_0^2-2\sigma_1^2}{2\sigma_0^2\sigma_1^2}y^2+\frac{3}{\sigma_1^2}y-\frac{3}{2\sigma_1^2}\right)dy+\kappa_1\\
	&\stackrel{(c)}{=}& \kappa_2,\label{third-abs2}
\end{IEEEeqnarray}
where $\kappa_1$ and $\kappa_2$ are positive constants. Here, $(a)$ follows because $\mathbb{E}_{Q_0}\left[\mathcal{T}_{\text{test}}(Y)\right]=0$, $(b)$ follows because the function $\sigma\mapsto \frac{1}{\sigma}\exp\left(-\frac{(y-1)^2}{2\sigma^2}\right)$ is an increasing function of $\sigma$ as $y\to\infty$ and also because $\sigma_1^2\geq \sigma_2^2$, $(c)$ follows because $1.5\sigma_0^2\geq \frac{4}{3}\sigma_0^2\geq \sigma_1^2$ and thus the integral converges to a finite positive real number. 

Next, we analyze the second term of \eqref{third-abs1}:
\begin{IEEEeqnarray}{rCl}
	\mathbb{E}_{\tilde{Q}_1}\left[ \big|\mathcal{T}_{\test}(Y) -\mathbb{E}_{\tilde{Q}_1} [ \mathcal{T}_{\test}(Y) ] \big|^3 \right] &\leq & \mathbb{E}_{\tilde{Q}_1}\left[ \big|\mathcal{T}_{\test}(Y)\big|^3\right]+3\mathbb{E}_{\tilde{Q}_1}\left[ \big|\mathcal{T}_{\test}(Y)\big|^2\right]\cdot \big| \mathbb{E}_{\tilde{Q}_1} [ \mathcal{T}_{\test}(Y) ]\big|\\
	&&\hspace{1cm}+3\mathbb{E}_{\tilde{Q}_1}\left[ \big|\mathcal{T}_{\test}(Y)\big|\right]\cdot \big| \mathbb{E}_{\tilde{Q}_1} [ \mathcal{T}_{\test}(Y) ]\big|^2+\big| \mathbb{E}_{\tilde{Q}_1} [ \mathcal{T}_{\test}(Y) ]\big|^3\\
	&\leq & \mathbb{E}_{\tilde{Q}_1}\left[ \big|\mathcal{T}_{\test}(Y)\big|^3\right]+3\mathbb{E}_{\tilde{Q}_1}\left[ \big|\mathcal{T}_{\test}(Y)\big|^3\right]^{\frac{2}{3}}\cdot \big| \mathbb{E}_{\tilde{Q}_1} [ \mathcal{T}_{\test}(Y) ]\big|\\
	&&\hspace{1cm}+3\mathbb{E}_{\tilde{Q}_1}\left[ \big|\mathcal{T}_{\test}(Y)\big|^3\right]^{\frac{1}{3}}\cdot \big| \mathbb{E}_{\tilde{Q}_1} [ \mathcal{T}_{\test}(Y) ]\big|^2+\big| \mathbb{E}_{\tilde{Q}_1} [ \mathcal{T}_{\test}(Y) ]\big|^3,\label{step500}
\end{IEEEeqnarray}
where inequality \eqref{step500} follows because $\mathbb{E}\left[|\mathcal{T}_{\test}|^2\right]\leq \mathbb{E}\left[|\mathcal{T}_{\test}|^3\right]^{\frac{2}{3}}$ and $\mathbb{E}\left[|\mathcal{T}_{\test}|\right]\leq \mathbb{E}\left[|\mathcal{T}_{\test}|^3\right]^{\frac{1}{3}}$. Thus, it remains to prove that $\mathbb{E}_{\tilde{Q}_1}\left[ \big|\mathcal{T}_{\test}(Y)\big|^3\right]$ and $\big| \mathbb{E}_{\tilde{Q}_1} [ \mathcal{T}_{\test}(Y) ]\big|$ are bounded. We first write $\mathbb{E}_{\tilde{Q}_1}\left[\mathcal{T}_{\text{test}}(Y)\right]$ as follows: 
\begin{IEEEeqnarray}{rCl}
	\mathbb{E}_{\tilde{Q}_1}\left[\mathcal{T}_{\text{test}}(Y)\right]&=&\mathbb{E}_{\tilde{Q}_1}\left[\frac{\tilde{Q}_1(Y)-\tilde{Q}_2(Y)}{Q_0(Y)}\right]\\
	&=&\mathbb{E}_{\tilde{Q}_1-Q_0}\left[\frac{\tilde{Q}_1(Y)-\tilde{Q}_2(Y)}{Q_0(Y)}\right]\\
	&=& \mathbb{E}_{\tilde{Q}_1-Q_0}\left[\frac{\tilde{Q}_1(Y)-Q_0(Y)-(\tilde{Q}_2(Y)-Q_0(Y))}{Q_0(Y)}\right]\\
	&=& \chi_{\text{G},1}-\rho_{\text{G}}\\&=&\kappa_3, \label{third-abs3}
\end{IEEEeqnarray}
where $\kappa_3$ is a positive constant.
Now, it remains to show that $\mathbb{E}_{\tilde{Q}_1}\left[ \big|\mathcal{T}_{\test}(Y)\big|^3\right]$ is also bounded.
\begin{IEEEeqnarray}{rCl}
	\mathbb{E}_{\tilde{Q}_1}\left[ \big|\mathcal{T}_{\test}(Y)  \big|^3 \right] 
	&=& \mathbb{E}_{\tilde{Q}_1}\left[ \Bigg| \frac{\tilde{Q}_1(Y)-\tilde{Q}_2(Y)}{Q_0(Y)}  \Bigg|^3 \right]\\
	&=& \frac{1}{\sqrt{2\pi}}\int_{-\infty}^{\infty} \Bigg|\frac{1}{\sigma_1}\exp\left(-\frac{(y-1)^2}{2\sigma_1^2}\right)-\frac{1}{\sigma_2}\exp\left(-\frac{(y-1)^2}{2\sigma_2^2}\right)\Bigg|^3\cdot \sigma_0^3\exp\left(\frac{3y^2}{2\sigma_0^2}\right)\cdot \frac{1}{\sigma_1}\exp\left(-\frac{(y-1)^2}{2\sigma_1^2}\right)  dy\nonumber\\
	&=& \frac{1}{\sqrt{2\pi}}\int_{-\infty}^{\infty} \frac{1}{\sigma_1^3}\exp\left(-\frac{3(y-1)^2}{2\sigma_1^2}\right)\cdot \sigma_0^3\exp\left(\frac{3y^2}{2\sigma_0^2}\right)\cdot \frac{1}{\sigma_1}\exp\left(-\frac{(y-1)^2}{2\sigma_1^2}\right)  dy\nonumber\\
	&\leq & \frac{\sigma_0^3}{\sqrt{2\pi}\sigma_1^4}\int_{-\infty}^{\infty}  \exp\left(\frac{3y^2}{2\sigma_0^2}\right)\cdot \exp\left(-\frac{2(y-1)^2}{\sigma_1^2}\right)  dy+\kappa_4\\
	&=&\frac{\sigma_0^3}{\sqrt{2\pi}\sigma_1^4}\int_{-\infty}^{\infty}   \exp\left(-y^2\left(\frac{2}{\sigma_1^2}-\frac{3}{2\sigma_0^2}\right)+y\frac{4}{\sigma_1^2}-\frac{2}{\sigma_1^2}\right)  dy+\kappa_4\\
	&=& \kappa_5,\label{third-abs4}
\end{IEEEeqnarray}
where $\kappa_4$ and $\kappa_5$ are positive constants. The last equality follows because $\sigma_1^2\leq \frac{4}{3}\sigma_0^2$ and so the integral has a finite value. Combining \eqref{third-abs1}, \eqref{third-abs2}, \eqref{step500}, \eqref{third-abs3} and \eqref{third-abs4}, we get:
\begin{IEEEeqnarray}{rCl}
	&&\sum_{i=1}^n \mathbb{E}_{W_1(\cdot|x_i^{*})}\left[ \big|\mathcal{T}_{\test}(Y_i) -\mathbb{E}_{W_1(\cdot|x_i^{*})} [ \mathcal{T}_{\test}(Y_i) ] \big|^3 \right]\\
	&&\hspace{1cm}=n(1-\mu_{1,n}) \mathbb{E}_{Q_0}\left[ \big|\mathcal{T}_{\test}(Y) -\mathbb{E}_{Q_0} [ \mathcal{T}_{\test}(Y) ] \big|^3 \right] +n\mu_{1,n} \mathbb{E}_{\tilde{Q}_1}\left[ \big|\mathcal{T}_{\test}(Y) -\mathbb{E}_{\tilde{Q}_1} [ \mathcal{T}_{\test}(Y) ] \big|^3 \right]\nonumber\\
	&&\hspace{1cm}\leq n(1-\mu_{1,n})\kappa_2+n\mu_{1,n}(\kappa_5+3\kappa_5^{2/3}\kappa_3+3\kappa_5^{1/3}\kappa_3^2+\kappa_3^2)\\
	&&\hspace{1cm}\leq n(\kappa_2+\kappa_5+3\kappa_5^{2/3}\kappa_3+3\kappa_5^{1/3}\kappa_3^2+\kappa_3^2)\\
	&&\hspace{1cm}\triangleq n\mathbf{T}.
	\end{IEEEeqnarray}
Thus, the above replaces \eqref{third-abs5} and the rest of the proof in Appendix~\ref{lem2-proof} holds.

\underline{\textit{Evaluation of the involved information quantities}}:

We now evaluate the information quantities involved in Theorem~\ref{opt-thm}.
First, consider the KL-divergence term $D(\tilde{Q}_s\|Q_0)$ as follows:
\begin{IEEEeqnarray}{rCl}
	D(\tilde{Q}_s\|Q_0) &=& -H_{\tilde{Q}_s}(Y)+\mathbb{E}_{\tilde{Q}_s}\left[\log\frac{1}{Q_0(Y)}\right]\\
	&=&-\frac{1}{2}\log\left( 2\pi e \sigma_s^2 \right)+\mathbb{E}_{\tilde{Q}_s}\left[\log\frac{1}{Q_0(Y)}\right]\\
		&=&-\frac{1}{2}\log\left( 2\pi e \sigma_s^2 \right)+\mathbb{E}_{\tilde{Q}_s}\left[\frac{1}{2}\log (2\pi e \sigma_0^2)+\frac{Y^2}{2\sigma_0^2}\log (e)\right]\\
		&=&\frac{1}{2}\log \frac{\sigma_0^2}{\sigma_s^2}+\frac{\log e}{2\sigma_0^2}\;\mathbb{E}_{\tilde{Q}_s}\left[Y^2\right]\\
		&=&\frac{1}{2}\log \frac{\sigma_0^2}{\sigma_s^2}+\frac{\log e}{2\sigma_0^2}\cdot (\sigma_s^2+1)\label{DG-ineq1}\\
		&=& D_{\text{G}}.\label{DG-def1}
	\end{IEEEeqnarray}
From \eqref{DG-ineq1} and the assumption $D(\tilde{Q}_1\|Q_0)=D(\tilde{Q}_2\|Q_0)$, we get the condition \eqref{Gaus-cons}.
Next, consider the chi-squared distance $\chi_2(\tilde{Q}_s\|Q_0)$ as follows:
\begin{IEEEeqnarray}{rCl}
	\chi_2\left(\tilde{Q}_s\|Q_0\right) &=& \mathbb{E}_{Q_0}\left[\left(\frac{\tilde{Q}_s(Y)}{Q_0(Y)}\right)^2\right]-1\\
	&=&\frac{\sigma_0^2}{\sigma_s^2}\;\mathbb{E}_{Q_0}\left[ \exp \left(-\frac{2(Y-1)^2}{2\sigma_s^2}+\frac{2Y^2}{2\sigma_0^2}\right) \right]-1\\
	&=&\frac{\sigma_0}{\sqrt{2\pi}\sigma_s^2}\;\int_{-\infty}^{\infty}\exp\left(-\frac{2(y-1)^2}{2\sigma_s^2}+\frac{y^2}{2\sigma_0^2}\right)dy-1\\
	&=&\frac{\sigma_0}{\sqrt{2\pi}\sigma_s^2}\;\int_{-\infty}^{\infty}\exp\left(-y^2\left(\frac{1}{\sigma_s^2}-\frac{1}{2\sigma_0^2}\right)+\frac{2}{\sigma_s^2}y-\frac{1}{\sigma_s^2}\right)dy-1 \\
	&=&\frac{\sigma_0}{\sqrt{2\pi}\sigma_s^2}\;\int_{-\infty}^{\infty}\exp\left(-y^2\left(\frac{1}{\sigma_s^2}-\frac{1}{2\sigma_0^2}\right)+\frac{2}{\sigma_s^2}y-\frac{1}{\sigma_s^2-\frac{\sigma_s^4}{2\sigma_0^2}}+\frac{1}{\sigma_s^2-\frac{\sigma_s^4}{2\sigma_0^2}}-\frac{1}{\sigma_s^2}\right)dy-1 \\
	&=&\frac{\sigma_0}{\sqrt{2\pi}\sigma_s^2}\;\exp\left(\frac{1}{\sigma_s^2}\left(\frac{1}{1-\frac{\sigma_s^2}{2\sigma_0^2}}-1\right)\right)\int_{-\infty}^{\infty} \exp \left( -\left(\sqrt{\frac{1}{\sigma_s^2}-\frac{1}{2\sigma_0^2}}y-\sqrt{\frac{1}{\sigma_s^2-\frac{\sigma_s^4}{2\sigma_0^2}}}\right)^2 \right)dy-1\\
	&=&\frac{\sigma_0}{\sqrt{2\pi}\sigma_s^2}\;\exp\left(\frac{1}{\sigma_s^2}\left(\frac{1}{1-\frac{\sigma_s^2}{2\sigma_0^2}}-1\right)\right)\frac{1}{\sqrt{\frac{1}{\sigma_s^2}-\frac{1}{2\sigma_0^2}}}\int_{-\infty}^{\infty} \exp \left( -u^2 \right)du-1\\
	&=&\frac{\sigma_0}{\sigma_s}\;\exp\left(\frac{1}{\sigma_s^2}\left(\frac{1}{1-\frac{\sigma_s^2}{2\sigma_0^2}}-1\right)\right)\frac{1}{\sqrt{2-\frac{\sigma_s^2}{\sigma_0^2}}}-1\\
	&=&\chi_{\text{G},s}.\label{chi-term}
	\end{IEEEeqnarray}
Finally, we evaluate the term $\rho(\tilde{Q}_1,\tilde{Q}_2,Q_0)$ as the following:
\begin{IEEEeqnarray}{rCl}
	\rho(\tilde{Q}_1,\tilde{Q}_2,Q_0) &=& \mathbb{E}_{Q_0}\left[\frac{\tilde{Q}_1(Y)}{Q_0(Y)}\cdot \frac{\tilde{Q}_2(Y)}{Q_0(Y)}\right]-1\\
	&=& \frac{\sigma_0}{\sqrt{2\pi}\sigma_1\sigma_2}\int_{-\infty}^{\infty}\exp\left(-\frac{(y-1)^2}{2\sigma_1^2}-\frac{(y-1)^2}{2\sigma_2^2}+\frac{y^2}{2\sigma_0^2}\right)dy-1\\
	&=&\frac{\sigma_0}{\sqrt{2\pi}\sigma_1\sigma_2}\int_{-\infty}^{\infty}\exp\left(-y^2\left(\frac{1}{2\sigma_1^2}+\frac{1}{2\sigma_2^2}-\frac{1}{2\sigma_0^2}\right)+y\left(\frac{1}{\sigma_1^2}+\frac{1}{\sigma_2^2}\right)-\left(\frac{1}{2\sigma_1^2}+\frac{1}{2\sigma_2^2}\right)\right)dy-1\\
	&=&\frac{\sigma_0}{\sqrt{2\pi}\sigma_1\sigma_2}\int_{-\infty}^{\infty}\exp\Bigg(-y^2\left(\frac{1}{2\sigma_1^2}+\frac{1}{2\sigma_2^2}-\frac{1}{2\sigma_0^2}\right)+y\left(\frac{1}{\sigma_1^2}+\frac{1}{\sigma_2^2}\right)-\frac{\left(\frac{1}{2\sigma_1^2}+\frac{1}{2\sigma_2^2}\right)^2}{\frac{1}{2\sigma_1^2}+\frac{1}{2\sigma_2^2}-\frac{1}{2\sigma_0^2}}\nonumber\\&&\hspace{4cm}+\frac{\left(\frac{1}{2\sigma_1^2}+\frac{1}{2\sigma_2^2}\right)^2}{\frac{1}{2\sigma_1^2}+\frac{1}{2\sigma_2^2}-\frac{1}{2\sigma_0^2}}-\left(\frac{1}{2\sigma_1^2}+\frac{1}{2\sigma_2^2}\right)\Bigg)dy-1\\
	&=&\frac{\sigma_0}{\sqrt{2\pi}\sigma_1\sigma_2}\exp\left( \frac{\left(\frac{1}{2\sigma_1^2}+\frac{1}{2\sigma_2^2}\right)\cdot \frac{1}{2\sigma_0^2}}{\frac{1}{2\sigma_1^2}+\frac{1}{2\sigma_2^2}-\frac{1}{2\sigma_0^2}} \right)\times\nonumber\\&&\hspace{3cm}\int_{-\infty}^{\infty} \exp\left(-\left(\sqrt{\frac{1}{2\sigma_1^2}+\frac{1}{2\sigma_2^2}-\frac{1}{2\sigma_0^2}}y-\frac{\frac{1}{2\sigma_1^2}+\frac{1}{2\sigma_2^2}}{\sqrt{\frac{1}{2\sigma_1^2}+\frac{1}{2\sigma_2^2}-\frac{1}{2\sigma_0^2}}}\right)^2\right)dy-1\\
	&=& \frac{\sigma_0}{\sqrt{2\pi}\sigma_1\sigma_2}\exp\left( \frac{\left(\frac{1}{2\sigma_1^2}+\frac{1}{2\sigma_2^2}\right)\cdot \frac{1}{2\sigma_0^2}}{\frac{1}{2\sigma_1^2}+\frac{1}{2\sigma_2^2}-\frac{1}{2\sigma_0^2}} \right)\cdot\frac{1}{\sqrt{\frac{1}{2\sigma_1^2}+\frac{1}{2\sigma_2^2}-\frac{1}{2\sigma_0^2}}}\int_{-\infty}^{\infty}\exp\left(-u^2\right)du-1\\
	&=& \frac{\sigma_0}{\sigma_1\sigma_2}\exp\left( \frac{\left(\frac{1}{\sigma_1^2}+\frac{1}{\sigma_2^2}\right)\cdot \frac{1}{2\sigma_0^2}}{\frac{1}{\sigma_1^2}+\frac{1}{\sigma_2^2}-\frac{1}{\sigma_0^2}} \right)\cdot\frac{1}{\sqrt{\frac{1}{\sigma_1^2}+\frac{1}{\sigma_2^2}-\frac{1}{\sigma_0^2}}}-1\\
	&=&\rho_{\text{G}}.\label{rho-term}
	\end{IEEEeqnarray}
Moreover, consider the fact that 
\begin{IEEEeqnarray}{rCl}
	\Delta = \chi_2(\tilde{Q}_1\|Q_0)+\chi_2(\tilde{Q}_2\|Q_0)-2\rho(\tilde{Q}_1,\tilde{Q}_2,Q_0)=\Delta_{\text{G}}.\label{DeltaG}
	\end{IEEEeqnarray}

The proof is concluded by combining \eqref{DG-def1} with \eqref{DeltaG} and \eqref{chi-term} with \eqref{rho-term}.

\bibliographystyle{IEEEtran}
\bibliography{references}

\end{document}